\documentclass[usenatbib]{mn2e}
\usepackage{amsmath,astrojournals,fleqn,epsfig,txfonts}

\catcode`\@=11
\def\gta{\ifmmode{\,\mathrel{\mathpalette\@versim>\,}}
    \else{$\,\mathrel{\mathpalette\@versim>}\,$}\fi}
\def\lta{\ifmmode{\,\mathrel{\mathpalette\@versim<\,}}
    \else{$\,\mathrel{\mathpalette\@versim<}\,$}\fi}
\def\@versim#1#2{\lower 2.9truept \vbox{\baselineskip 0pt \lineskip
    0.5truept \ialign{$\m@th#1\hfil##\hfil$\crcr#2\crcr\sim\crcr}}}
\catcode`\@=12  

\def\kms{\,{\rm km}\,{\rm s}^{-1}}
\def\mas{\,{\rm mas}}
\def\masyr{\,{\rm mas\,yr^{-1}}}
\def\mag{\,{\rm mag}}

\def\Gyr{\,{\rm Gyr}}
 
\def\pc{\,{\rm pc}}
\def\kpc{\,{\rm kpc}}

\def\feh{\hbox{[Fe/H]}}
\def\feha{{[Fe/H]_{DR7}}}
\def\fehC{{[Fe/H]_{Car}}}
\def\sigra{\sigma_{\rm RA}}
\def\sigdec{\sigma_{\rm DEC}}

\def\dex{\,{\rm dex}}
\def\llg{\log({\rm g})}
\def\figref#1{Fig.~\ref{#1}}
\newcommand {\B}[1]{{\boldsymbol{#1}}}
\newcommand {\vsun}{{\B{\upsilon_\odot}}}
\newcommand {\Usun}{{U_{\!\odot}}}
\newcommand {\Wsun}{{W_{\!\odot}}}
\newcommand {\Vsun}{{V_{\!\odot}}}
\newcommand {\Rsun}{{R_{G,\odot}}}
 
\newcommand{\beq}{\begin{equation}}
\newcommand{\eeq}{\end{equation}}

\title[On the alleged halo duality]
{On the alleged duality of the Galactic halo}

\author[R. Sch\"onrich et al.]{Ralph Sch\"onrich\thanks{E-mail: rasch@mpa-garching.mpg.de}, Martin Asplund, Luca Casagrande\\
 Max Planck Institute for Astrophysics, Garching, Germany \\ }

\begin{document}

\date{Draft, December 2, 2010}

\pagerange{\pageref{firstpage}--\pageref{lastpage}} \pubyear{2010}

\maketitle

\label{firstpage}

\begin{abstract}

We examine the kinematics of the Galactic halo based on SDSS/SEGUE
data by Carollo et al. (2007, 2010). We find that their claims of a
counter-rotating halo are the result of substantial biases
in distance estimates (of order $50\%$): the claimed retrograde component, which makes up only a tiny fraction of the entire sample, prone to contaminations, is identified as the tail of distance overestimates. The
strong overestimates also result in a lift in the vertical
velocity component, which explains the large altitudes those objects
were claimed to reach. Errors are worst for the lowest metallicity
stars, which explains the metal-poor nature of the artificial component. We also argue that measurement errors were not properly accounted for and that the use of Gaussian fitting on intrinsically non-Gaussian Galactic components invokes the identification of components that are distorted or even artificial. Our evaluation of the data leads to a revision of the estimated velocity ellipsoids and does not yield any reliable evidence for a
counterrotating halo component. If a distinct counterrotating halo component exists it
must be far weaker than claimed by Carollo et al. Finally we note that their revised analysis presented in Beers et al. (2011) does not alleviate our main concerns.

\end{abstract}

\begin{keywords}
galaxies: haloes - stars: distances - Galaxy: solar neighbourhood - Galaxy: halo -  Galaxy: kinematics and dynamics
- Galaxy: structure
\end{keywords} 
\newcounter{mytempeqncnt}
\section{Introduction}

Galactic haloes are an excellent testbed for cosmology and galactic
dynamics. Their exploration can constrain the early assembly of galaxies
as well as the dynamics of accretion of smaller galaxies. Our Milky
Way offers an ideal case for those investigations, as we can directly
obtain the detailed parameters like kinematics, elemental abundances
and physical properties of single stars surrounding us. New material
is still being accreted into the Galactic halo, as the
numerous streams and newly discovered dwarf galaxies confirm \citep[see
  e.g.][]{Ibata95, Klement06, Belokurov07}. As metallicity gradients go in lockstep with star formation, young accreted objects may be more metal-poor than old stars from the inner Galaxy and
thus metallicity can not be simply used as a cosmic clock. This would
also make it plausible that a later accreted halo component could
indeed be on average younger and more metal poor than the older
parts. The formation of at least parts of the halo by
accretion (combined with later adiabatic contraction) could give rise to differences between early more turbulent accretion/collapse processes 
and later accretion, which might leave an imprint in differences
between the inner and outer halo \citep[see e.g.][]{Cooper10}. Another possible source of
discrepancies between inner and outer halo is dynamical friction, which
could be more efficient for prograde than for retrograde infall
\citep{Quinn86, Byrd86}. This could give rise to a different rotational
signature for accreted material in the outer Galactic halo compared to
the inner regions \citep{Murante10}. 

Historically (and as well today), halo stars have been extremely difficult to identify,
particularly in local samples, e.g. demonstrated by the historic
argument between Oort and Str\"omberg \citep[][]{Oort26, Strom27}. Like it is
practically impossible to get a clean selection into thin and thick
Galactic disc based on kinematics \citep[][]{SB09b}, we face the analogous problem between
thick disc and the prograde stars of the halo. Wrong assumptions about the kinematics of the Galactic disc will thus affect results on the halo component. Soon after the existence of the Galactic halo was established \citep[][]{Schwarzschild52, Eggen62}, the central question was raised if the stars of the inner and the outer
halo had the same properties or if gradients or even breaks in
metallicities or kinematics existed with galactocentric radius. Two main strategies to identify and examine halo stars have been used in the past: either stars in the solar neighbourhood are studied,
classified according to their kinematics and then conclusions about
the structure further away are drawn by extrapolation
\citep[e.g.][]{jesper}, or the surveys concentrate on bright objects in
the outer halo regions, such as RR Lyrae variables or globular
clusters \citep[e.g.][]{Sandage70}. 
The second alternative allows to directly map the spatial structure by those standard candles with good distance information \citep[e.g.][]{Saha85}. This strategy implies selection biases: for example the position and presence of RR Lyrae stars on the horizontal branch are correlated with metallicity and age, while it is not known if the formation of globular clusters is representative also for all halo field stars.

Claims of differences between the inner and outer Galactic halo are
almost as old as the discovery of the halo itself. After
\cite{Bergh67} discussed differences in metallicity and the second
parameter between the haloes of the Milky Way and those of its neighbouring galaxies (M31, M33), \cite{Searle78} found that Galactic clusters in the
outer regions showed a larger scatter in the ratio of blue to red
horizontal branch stars than inner halo globular clusters, which they
interpreted as a signature of an age spread. \cite{Preston90} found a
similar difference in field BHB stars.\footnote{While it is clear that
  age is one parameter which will cause an older globular cluster to
  be bluer than a younger one at the same metallicity, whether this is
  the dominant cause of differences in horizontal branch morphology is
  still debated (see, for example, \citet{Dotter10} and
  \cite{VandenBerg00} for opposing views.)}  Differences in kinematics
have also been suggested between inner and outer halo globular
clusters \citep{zinn93}, although precision and reliability of estimates in this respect are limited by the small number of available globular clusters. Various claims of an asymmetry in the halo azimuthal velocity
distribution with an extended tail to retrograde orbits  have been made \citep[e.g.][]{Norris89}. \cite{Majewski92} even found the entire halo to be on average counterrotating,
could, however, not find any significant velocity gradient. \cite{Ryan92} pointed out that measurements of kinematics based on proper motions were particularly vulnerable to distance errors and showed that
overestimated distances for halo stars can lead to false
identifications of counter-rotating stars. 

In this paper we will revisit the recent claim by \cite{Carollo07} (hereafter C07) and \cite{Carollo10} (hereafter C10) that the Galactic halo consists of two components: a more metal-poor counterrotating component with larger scaleheights, distinct from a slightly prograde component and starts dominating the halo at high altitudes in their analysis.\footnote{This dominance of a retrograde component in the outer halo has recently been contested by \cite{Deason10} although they find a retrograde motion for metal-poor ($\feh < -2$) and prograde ($\feh > -2$) for metal-rich stars using a BHB sample in the outer Galactic halo. In our view, this issue needs to be further investigated, as they assume constant $g$ band magnitudes for the horizontal branch in a region affected by the blue tail, which spans of order $2 \mag$ in $g$ band luminosity. If a considerable fraction of the halo giants is in the blue tail, their colour and temperature cuts remove a large part of this tail, but still leave $BHB$ members spanning $\sim 0.7$ magnitudes, as we tested it on SDSS photometry of metal-poor globular clusters known to have such a strong blue tail ($M3$, $M13$, $M15$ and $M92$) and the BASTI isochrones.} In particular we will carefully re-examine their distance estimation procedure; we will focus on C10 as this paper deviates from its precursor mostly by the larger sample size. To avoid relying on any of the uncertain available distance calibrations, we apply in parallel both the C10 distances and two native SDSS main sequence distance calibrations, checking results additionally with an isochrone method. In Section
\ref{sec:segue} we outline those methods, discuss the SDSS/SEGUE data used for this
purpose and describe how the sample cleaning was performed. 

 Thereafter (Section \ref{sec:grav}) we discuss the underlying assumptions and the
reliability of gravity estimates used to sort stars into different
sequences as well as the actual assumptions for absolute magnitudes by
C07 and C10. We will show that their claim to have distances precise to $10-20 \%$ is unsupported and that the C10 sample contains a class of stars with significant distance overestimates by being sorted into unphysical positions in the HR diagram. In Section \ref{sec:sign} we present statistical
proofs of distance biases in the sample and in Section \ref{sec:vel} we discuss the implications of different distance schemes on kinematics and the inner-outer halo dissection.

\section{The calibration sample of SDSS and the Carollo dataset}
\label{sec:segue}

All the data used in this paper come from the Sloan Digital Sky Survey
\citep[SDSS,][]{york00}, and consist of spectroscopic
observations of stars from both SDSS-I and II and from the
SDSS-II/SEGUE survey \citep{Yanny09}. Stellar parameters for the stars
were estimated using the SEGUE Stellar Parameters Pipeline
\citep[SSPP,][]{Lee08a,Lee08b,carlos}.

For this study we use the calibration star sample from SDSS public data release $7$ (DR7).\footnote{http://casjobs.sdss.org/dr7/en/} As the colour transformations and distances used in C07 and
C10 are not part of the public data releases, we draw this information from the sample used by C10, which is a cleaned version of DR7 and which was kindly provided on our request.

The calibration star sample comprises two datasets, namely the
photometric calibration star sample and the reddening calibration star
sample. Querying the DR7 catalogue for these stars produces a total of
$42841$ entries, but many refer to identical objects, so that the
actual number of unique objects in the database is $33023$. The
interested reader is referred to the Appendix where we describe how
the cleaning of the sample from questionable objects has been performed that cuts down the sample to $28844$ stars.

Throughout the paper we make use of two different classes of distance determinations (see Appendix for details): On the one hand the distances used by C07 and C10 and on the other hand two derivations adopted from \cite{Ivz08}. When using the distances by C07 and C10 we examine the effects of their different sequences by both using their entire sample ("Carollo all") and using exclusively their dwarf stars ("Carollo dwarfs"). For the native SDSS calibrations we restrict ourselves to the dwarf stars, imposing in general a gravity limit of $log(g) > 4.1$ to reduce the impact of giant and subgiant contamination). We there use two different schemes: An adopted main sequence derivation, where we increase distances defined via their Eq. (A2) and (A3) by accounting for alpha enhancement and additionally decreasing the adopted absolute magnitudes of all stars by $0.1 \mag$, hereafter termed the adopted main sequence calibration, short "IvzMS"). Second the calibration favoured by \cite{Ivz08} using Eq. (A7) from their appendix, which is steepened towards the isochrones in order to account for age dependent effects, hereafter called the age dependent calibration (short: "IvzA7"). We would like to point out that we believe neither distance calibration to deliver the full truth, yet they are currently the most commonly used schemes and can give hints on the intrinsic uncertainties of all methods. To have an additional test we cross-checked and confirmed our findings with distances derived directly from using $12.5 \Gyr$ BASTI isochrones \citep[][]{Piet04, Piet06}; some details are provided in the Appendix.

C10 applied two different geometric cuts. The first was a
distance cut at an estimated distance of $4 \kpc$, limiting the errors on
velocities caused by proper motion errors. The second was a cut that
removed stars differing from the Sun by more than $1.5 \kpc$ in
galactocentric radius (i.e. outside $7.0 \kpc \le R_G \le 10 \kpc$
with their value of $\Rsun = 8.5 \kpc$ for the Sun). The latter cut was
reasoned in C10 by the use of their applied orbit model, which was
adopted from \cite{Chiba00} using a St\"ackel-type potential from
\cite{jesper}. 
As we do not make use of this orbit calculation, in this work we do not apply the
second cut, which removes of order one third of the
available stars. 

Due to the magnitude
ranges in the SEGUE survey the distance cuts imposed by C10 and C07 remove almost all giants from the
sample, a minor fraction of subgiants and very few dwarf stars. 
 The effects of both selections depend strongly on the adopted distances and thus the
absolute magnitudes of the stars. Applying the C10 distances to the sample, $21600$ stars remain within the distance
limit of $4 \kpc$, of which $14763$ have surface gravities $\llg
> 4.0$ and are thus classified as dwarf stars. With the adopted main sequence calibration $15808$
stars with $\llg > 4.0$ are found in the sample, which gives a first
hint to the more stretched distance scale by C10. 

Metallicities are taken from the DR7 pipeline adopted
values \citep[][]{Lee08a, Lee08b}. For a discussion of the different metallicity scales of DR7 and C10 the reader is referred to the Appendix. There we also describe in more detail how kinematics are derived from distances, radial velocities and proper motions.

\begin{figure}
\epsfig{file=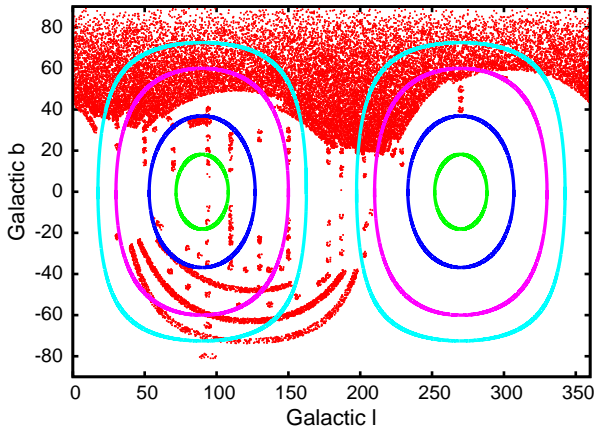,angle=-90,width=\hsize}
\caption{The distribution of all reliable stars in the
  calibration sample, fulfilling the temperature range limit and with
  acceptable kinematic information. To demonstrate the weak line-of-sight motion support
  of azimuthal velocities we plot the $0.95$ (innermost lenses, green),
  $0.8$, $0.5$ and $0.3$ (outermost lenses, light blue),
  contours of the angle term $\eta_V$ in Eq.(\ref{eq:nuv}) that
  quantifies the relative support by the direct line-of-sight velocity
  measurements. This demonstrates
  the vulnerability of $V$ velocities in this sample to distance
  errors as they cannot be found directly from line of sight 
  velocities.}\label{fig:geo}
\end{figure}

\section{Assessing the distance calibrations}\label{sec:grav}

The core assumptions in C10 are those about stellar
distances. They claim that by using the $\llg$ estimates from the
SEGUE stellar parameter pipeline stars can be reliably sorted into
clean sequences, i.e. main sequence, turn-off  and
subgiant/giant. The attraction of this idea, laid out in
\cite{Beers00}, is that it seems to reduce the distance errors to
simple uncertainties in colour and metallicity on well-determined
sequences. However, as we will see below, things are more complicated. 

\subsection{Effects of distance errors}

Selecting a star into the wrong position in the colour-magnitude
diagram results in a faulty estimate of its absolute magnitude and
thus an erroneous distance. As they are the most common population we would naively expect the largest
contamination to be main sequence stars mistakenly addressed as
turn-off stars. These will be assumed to be far
brighter than they are, hence their distance will be overestimated, bringing
many of them, especially halo stars, falsely into the retrograde tail
of the velocity distribution. This effect happens via the translation
of proper motions into velocities and thus prevails for samples that
have low support by radial velocities (cf. Section \ref{sec:geo} of the Appendix). \figref{fig:geo} depicts the
locations of stars (red dots) in Galactic longitude (x-axis) and latitude (y-axis). The
SDSS/SEGUE sample is (due to the location of the telescope in the
northern hemisphere and the strategy to avoid the high extinction in
the Galactic plane) largely concentrated away from the plane and
towards the Galactic North Pole. Consequently it has almost no points
at directions that would have high support of azimuthal velocities by
the direct line-of-sight velocity measurements. The contours in the
plot encircle the regions of high line-of-sight velocity support of $V_h$ in the sky.\footnote{Throughout the paper we distinguish between the velocities in
the solar rest frame and coordinate system $(U_h, V_h, W_h)$ and
velocities in the Galactic rest frame and Galactic cylindric
coordinate system $(U,V,W)$. For a short discussion of those we refer
to the Appendix.} Within the ellipses, the fraction $\eta_V$ (Eq.(\ref{eq:nuv}) in the Appendix) of the
line-of-sight velocity going into the heliocentric azimuthal velocity $V_h$ is larger than
$0.95$ (smallest lenses), $0.8$, $0.5$ and $0.3$ (largest
lenses). They demonstrate how heavily any analysis of the azimuthal
velocities has to rely on the transverse velocity component and thus proper motions and distance estimates.

The effect of distance errors in this process is easily understood:
Think of driving a car past a field that has a
rabbit sitting on it. As the speed of the car is known, the fact that the rabbit rests on the lawn can be derived
by the car driver from its apparent angular speed - if the distance is right. If
the natural size of the rabbit is overestimated, so will be its
distance and to explain its angular motion one wrongly infers that it
moves opposite to the car's direction of motion. And vice versa a
distance underestimate drags the estimated rabbit velocity towards
that of the car, i.e. it is wrongly inferred that the rabbit moves in
the same direction as the car does. As the Sun moves with a velocity
of more than $200 \kms$ around the Galactic centre, for this sample a
$10\%$ distance overestimate implies that an average halo star in the sample will
be wrongly pushed by $\sim 20 \kms$ into retrograde motion, and a
larger distance error will entail a proportionately larger retrograde
motion. 

It should also be mentioned that azimuthal velocities are subject to
an error similar to the \cite{Lutz73} bias: Even if there is no net bias
on estimated absolute magnitudes and thus of distances, a symmetric
magnitude error distribution will cause an asymmetric distribution of estimated distances with a longer tail in the overestimates. As these directly translate to the transverse velocities one will with any magnitude based distance scheme encounter asymmetric velocity errors that give rise to an extended counterrotating tail in the measured  halo azimuthal velocity distribution. This may also be related to the asymmetry found by \cite{Norris89}.

\subsection{Gravities}

\begin{figure}
\epsfig{file=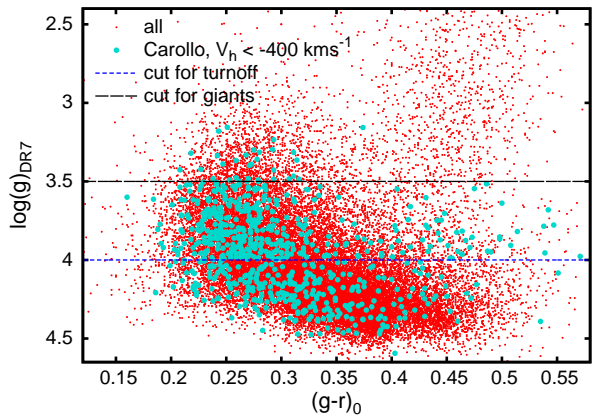,angle=-90,width=\hsize}
\caption{The distribution of the ``well behaved'' stars in the
  calibration sample (red crosses, without distance cut to make the giants visible) in the dereddened colour-gravity (DR7 pipeline)
  plane. Horizontal lines mark the cuts adopted by Carollo et
  al. between the dwarf, turn-off and subgiant/giant regions. Stars that have likely membership in their counter-rotating
  halo, i.e. with $V_{h} < -400 \kms$ and fulfilling $d < 4 \kpc$ are marked with blue points.}\label{fig:cg}
\end{figure}

~\figref{fig:cg} shows the positions of all cleaned sample stars in the colour -
$\llg$ plane (red points). The horizontal lines show the selection
regions in $\llg$ used by C10. Stars with $\llg > 4.0$
are classified as main sequence stars. Those with $\llg < 3.5$ are assumed to be subgiants or giants, while the
intermediate objects are classified into their "turn-off" branch. The
blue dots mark the objects in the C10 sample that have heliocentric
azimuthal velocities, $V_h$, smaller than $-400 \kms$, i.e. they are on highly
retrograde orbits and should be in the majority counter-rotating halo
objects according to C10. There can be
minor differences between the gravities directly from DR7 and
their sample, as we we use the best determined values when their are double or multiple entries for a star, while their $\llg$ values (which we do not have access to) are probably averaged.

The plot reveals one crucial problem with those criteria:  
although it may be expected that some
distinct branches of stars are present, they are not reflected by the SEGUE
measurements. While the upper giant branch is apparent, we cannot make
out a substantial decrease of densities between the main sequence
and subgiant regions. Measurement errors prevail, especially in
the turn-off region, and veto against a clean selection of the components. 

One possibility is that the measurement errors on $\llg$ are so large that it is not possible to
use the $\llg$ measurements to classify stars into these three
categories.  Also the true main sequence is inclined in gravity
versus colour. In this perspective the constant (colour and
metallicity independent) gravity cuts
applied by C07 and C10 do not appear
well founded. If one aims to select a pure dwarf sample, the
tightening of the constant cut relative to the inclined main
sequence, is, however, beneficial in reducing the turn-off-contamination.

The accuracy of the DR7 $\llg$ values is discussed by \citet{Lee08b},
who show estimated surface gravities for open and globular cluster stars
observed by SEGUE for the purpose of calibration of the
survey. As demonstrated by \citet{Lee08b} (see their Fig. 15 ff.) the low
spectral resolution results in a significant scatter in $\llg$ values: some turn-off stars have $\llg <
3.5$, and significant numbers of stars with $\llg$ measurements in the
region assigned by Carollo et al. to the turn-off ($3.5 < \llg < 4$)
are clearly either subgiants or main sequence stars. Thus we have to
expect a significant number of main sequence objects in the
turn-off band of the \cite{Carollo07, Carollo10} samples.

The consequent bias in distance estimation discussed above will
give an artificially enhanced fraction of retrograde halo stars
residing in the turn-off/subgiant regime. 
Although \figref{fig:cg} should play this effect
down with the red points being drawn from the main sample without
distance cuts (to keep the red giant branch visible), while the blue
circles satisfy $d < 4 \kpc$ in addition to the velocity cut, this crowding of the counterrotating
halo stars into the designated turn-off region is still prominent. The
distance cut is responsible for the missing enhancement in the giants.

\begin{figure}
\epsfig{file=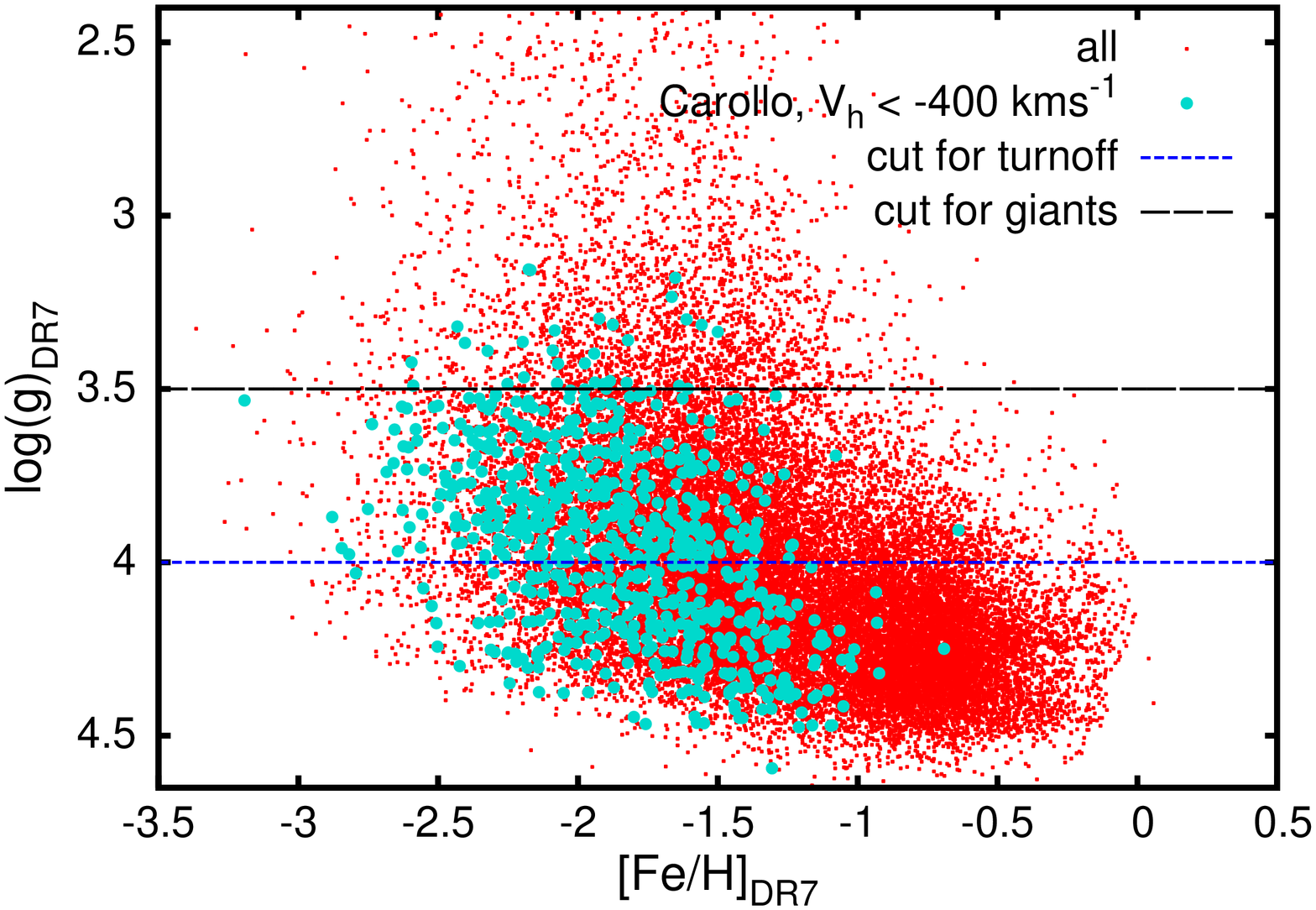,angle=-90,width=\hsize}
\epsfig{file=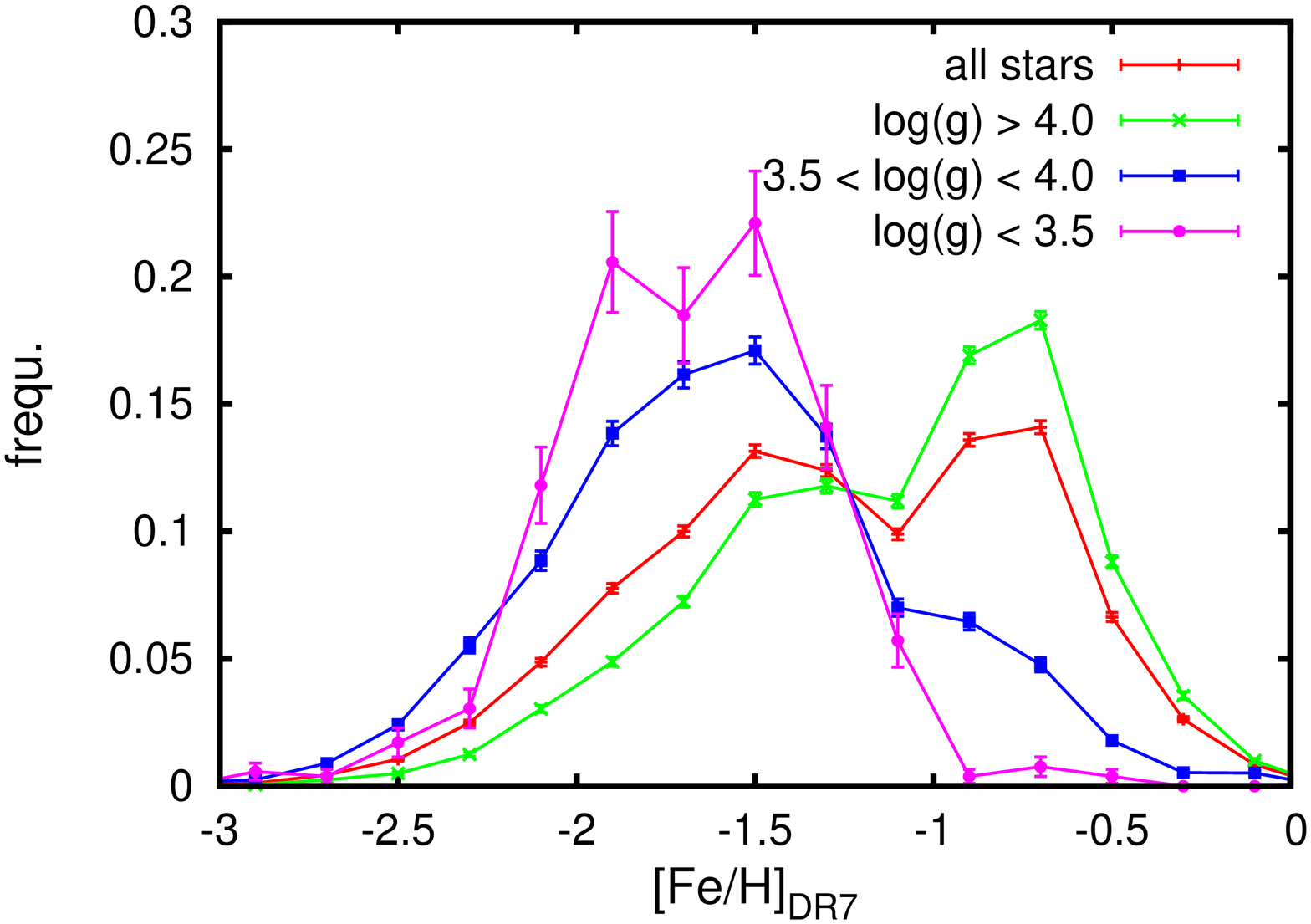,angle=-90,width=\hsize}
\epsfig{file=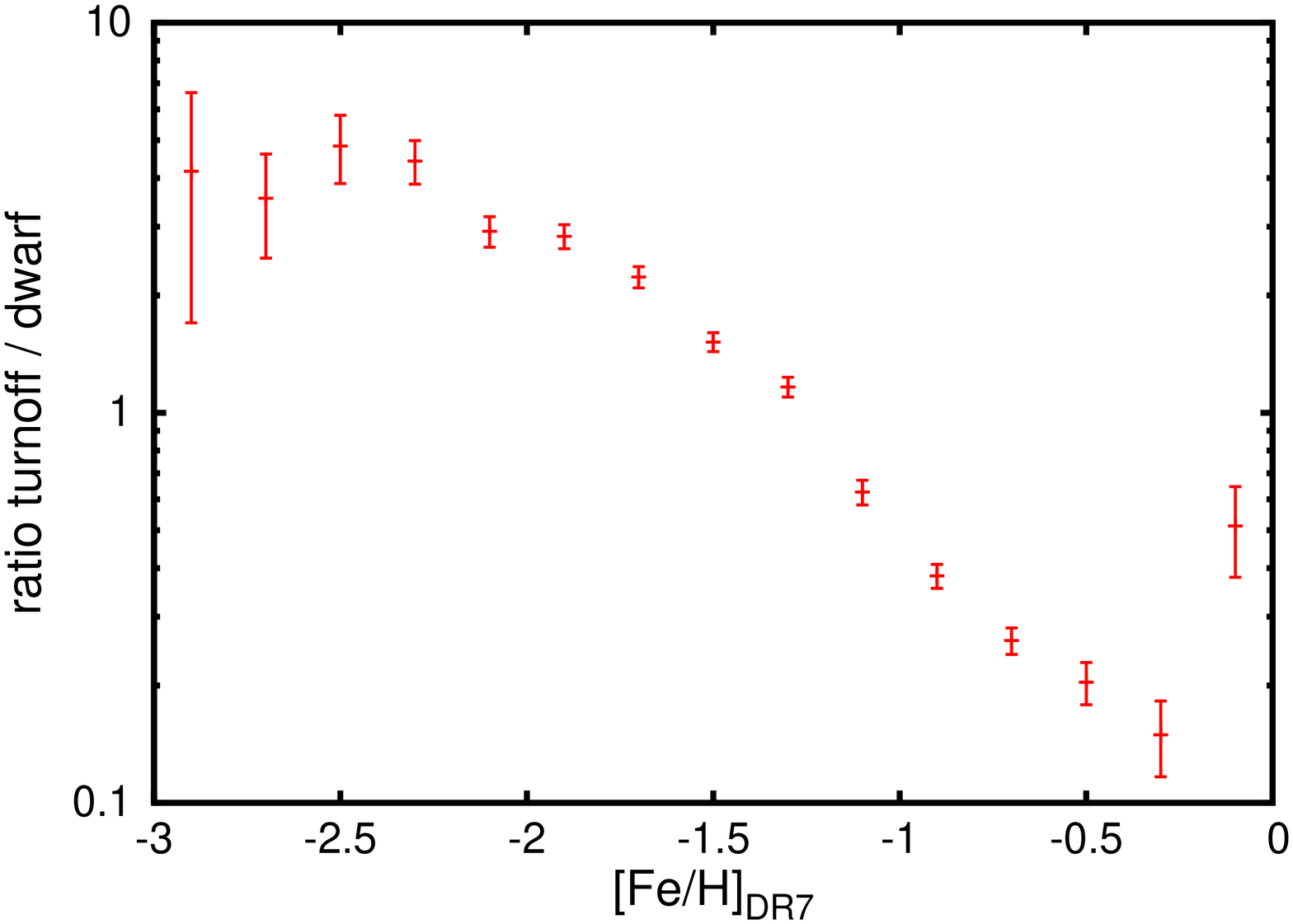,angle=-90,width=\hsize}
\caption{Top panel: The distribution of the cleaned "all star" sample in the metallicity- surface gravity plane. Horizontal lines mark the cuts adopted by
  Carollo et al. between the dwarf, "turn-off" and subgiant/giant
  regions. Stars in the C10 subsample displaying a clear membership to what might be a
  counter-rotating halo, i.e. with $V_{h} < -400.0 \kms$ are
  marked with blue points. Middle: Metallicity distributions for the
  different gravity classes used by Carollo et al., showing how low
  metallicities dominate the "turn-off" and "subgiant" stars. Bottom
  panel: Ratio of identified "turn-off" stars to dwarf stars as
  function of $\feha$ using the gravity cuts by Carollo et al. Even
  between $\feha \sim -1.5$ and $\feha \sim -2.5$ there is a strong uptrend.}
\label{fig:felg}
\end{figure}

Carollo et al. also find that their outer counterrotating halo members
display significantly lower metallicities than the average of the
inner halo. Could this be related to problems with
distance estimation as well? It would not be unreasonable to expect
the accuracy of the $\llg$ estimates to decrease for the stars of
lowest metallicity, increasing by this the fraction of fake turn-off
stars and subgiants, since the stellar lines become weaker. In fact, Ma
et al. (in prep.) find this effect. A better
understanding of why the ``counter-rotating'' component of Carollo et
al. has lower metallicity can be achieved from ~\figref{fig:felg}. In its top panel we plot the gravities against the
metallicity, again with red points for the entire sample and with blue
circles for the strongly retrograde stars. Stars with lower metallicities in the sample get on average
assigned lower gravities. This could arise from the fact that
determining gravities gets more difficult on the lowest
metallicity side. The inclination of the sample in the
metallicity-gravity plane favours distance overestimates for metal
poor stars, thus lowering the average metallicity of the
``counter-rotating'' component. This shift in classification of stars
is demonstrated in the middle panel of \figref{fig:felg}: Almost all
identified subgiants are metal-poor, while the average metallicity
rises towards the higher surface gravity categories. 
The bottom panel shows the ratio of the number of turn-off to main sequence
stars, $\rho$, against metallicity. 
One might argue that the pronounced rise of $\rho$ towards lower
metallicities be caused by the intrinsic rarity of metal-poor
objects. As the sample is dominated by magnitude cuts, the
different ratios could, however, just be explained by different geometry of the
subpopulations, which indeed puts stars with higher scaleheights
(e.g. halo) to more remote positions than those with low scaleheights
(e.g. disc). Since the C10 sample extends by definition less than $4 \kpc$
away from the plane it is hardly possible to explain how $\rho$ can
rise by a factor of $\sim 3$ from $\feha \sim -1.5$ to $ \feha \sim
-2.5$ as the large scaleheight of the halo should veto against
relative density variations of the halo populations by this amount.

\begin{figure}
\epsfig{file=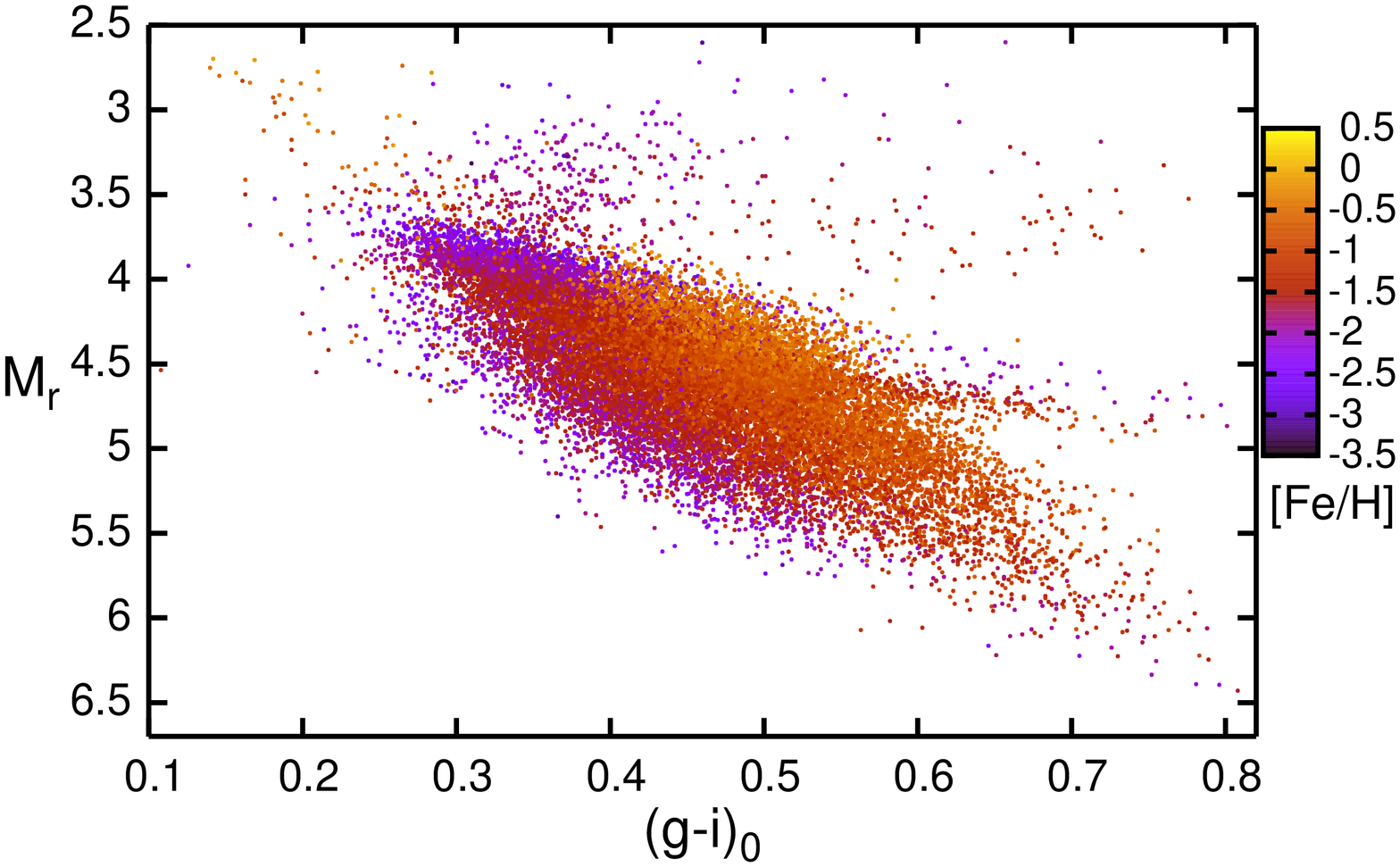,angle=-90,trim=15mm 10mm 6mm 10mm,clip, width=\hsize}
\epsfig{file=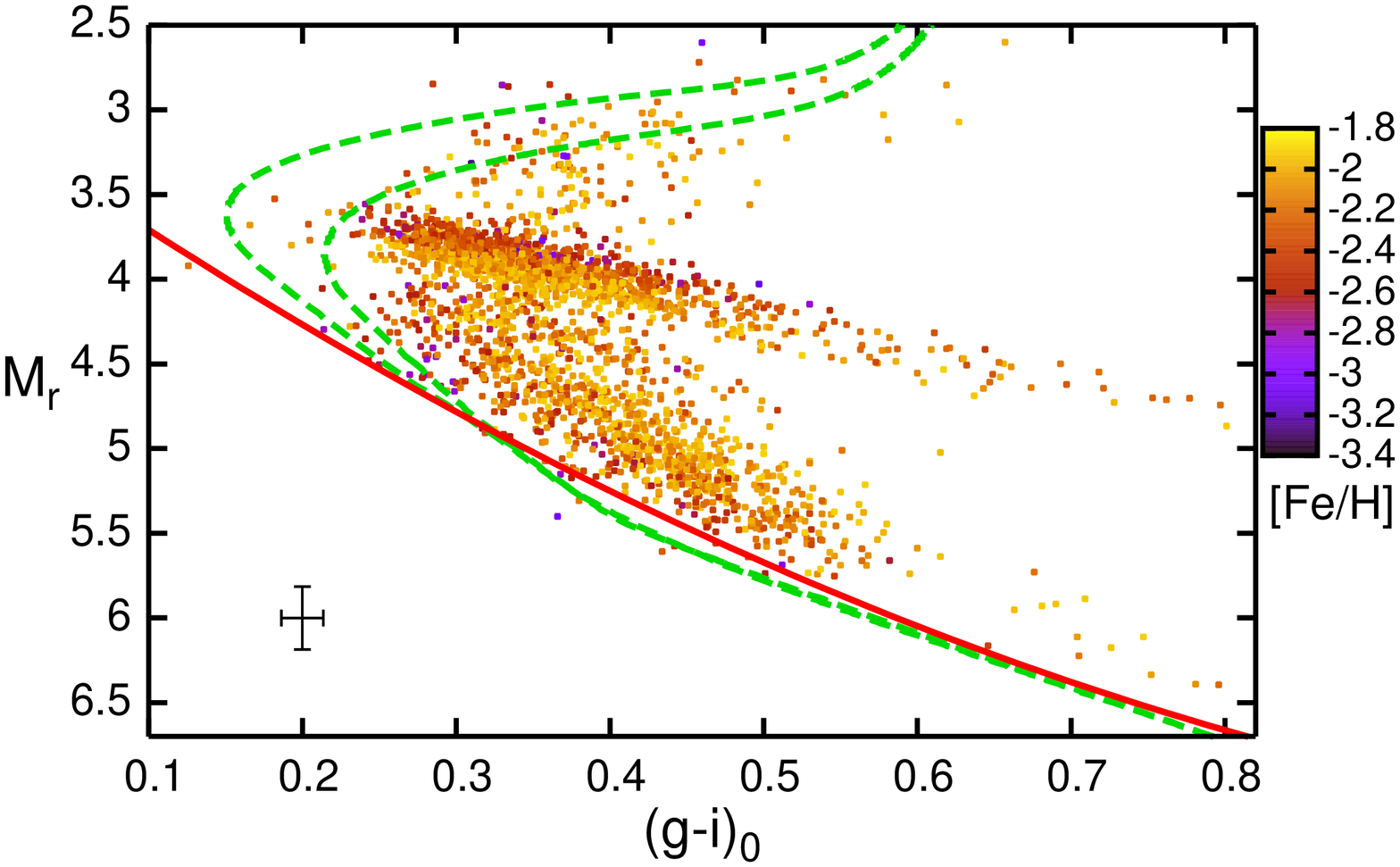,angle=-90,trim=15mm 10mm 6mm 10mm,clip,width=\hsize}
\epsfig{file=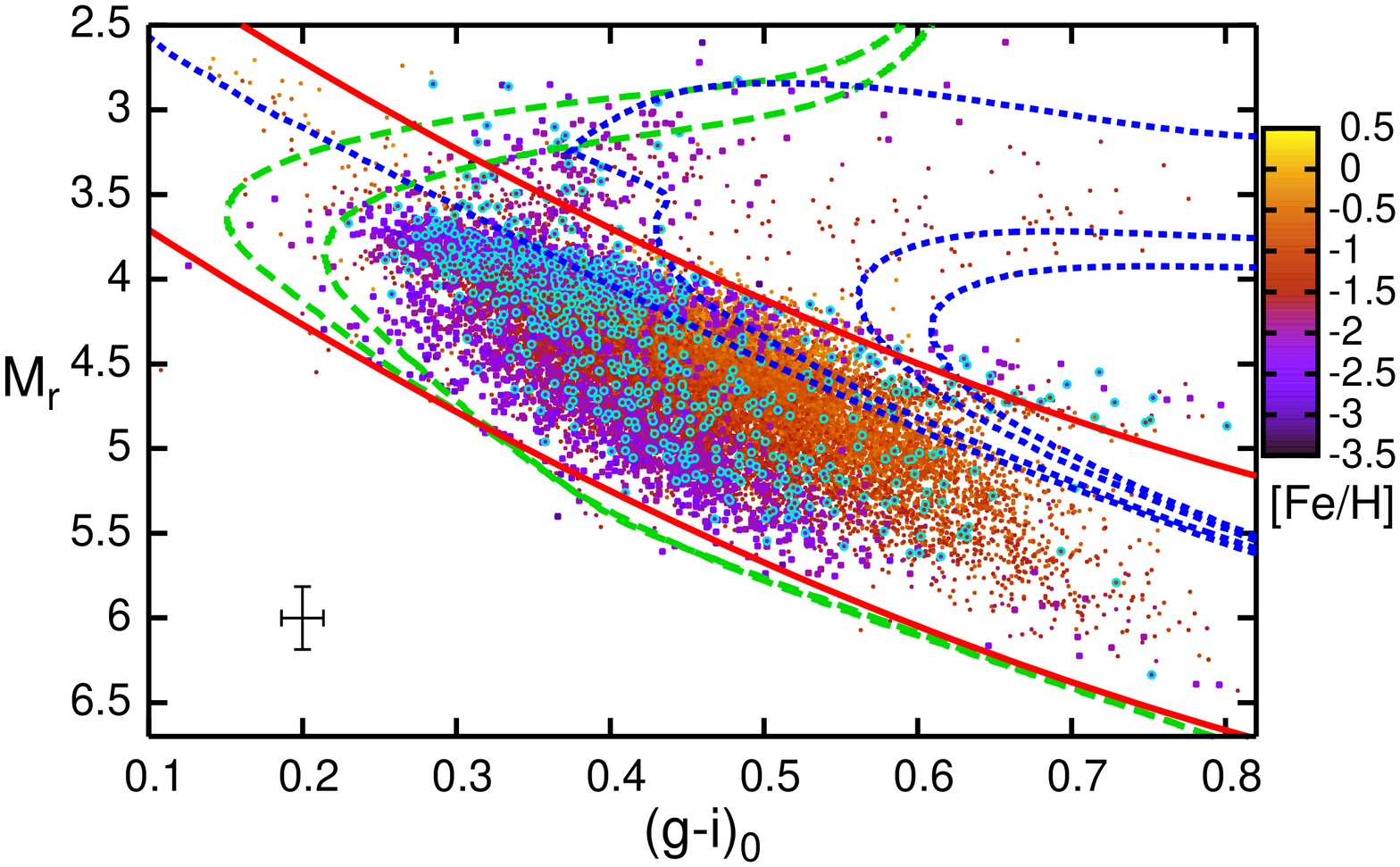,angle=-90,trim= 15mm 10mm 6mm 10mm,clip,width=\hsize}
\caption{
The distribution of the sample stars in the $(g-i)_0, M_r$
 plane with tiny crosses. Colours give the metallicity $\fehC$
 according to the colour bar on the right. The top panel contains all stars in the C10 sample that pass our quality cuts, while the middle
panel exclusively contains stars with $\feha < -1.9$. For
comparison we give the BASTI (Pietrinferni et al. 2004, 2006) alpha enhanced isochrones at
$\feh = -2.14$ of ages $10$ and $12.5$ Gyrs (green
lines) and the Ivezi{\'c} et al.(2008) absolute magnitude calibration
at the same iron abundance (red line). In the bottom panel the metal-poor
stars have been enhanced in size and stars with $V_{h} < -400 \kms$
derived with C10 distances are marked by light blue circles. Red lines
show the adopted main sequence calibration (IvzMS) at $\feh = -2.14$ and
$\feh = 0$. Dark blue lines show the BASTI isochrones at solar
metallicities for $1$, $4$, $10$ and $12.5$ Gyrs, green lines same as
in the middle panel. The crosses mark the typical photometric errors and uncertainties in the isochrones.
}\label{fig:HRrgi}
\end{figure}

\subsection{Absolute magnitudes}\label{sec:HR}

Having discussed the surface gravity estimates it is time to turn
towards the actual assumptions on absolute magnitudes of stars from
which the distances are inferred. The claim by C10 to
reach an accuracy of ``$10-20 \%$'' in distance estimates is
predicated on the ability to cleanly select stars according
to their spectroscopically determined gravity into several
branches. If such a selection was feasible, it would indeed limit the uncertainties to
those caused by metallicity, reddening and photometric errors. The
reader should bear in mind that the low relative density of the outer
halo component found by Carollo et al. implies that even a
contamination on the $1 \%$ level (i.e. $200$ out of $\sim 20000$ stars)
can alter the results.  

The sorting into different branches via the formalism of Beers et
al. (2000) can be seen in the top panel of ~\figref{fig:HRrgi}. The
plot shows all stars in the C10 sample in the $(g-i)_0, M_r$
plane that do not have warning
flags. The absolute magnitudes were
derived directly from the C10 distances using the
distance modulus via
\begin{equation}
M_r = r_0 - 5\log_{10}(\frac{d_{Car}}{0.01 \kpc})
\end{equation}
where $M_r$ is the derived absolute magnitude, $r_0$ is the reddening
corrected apparent magnitude from the DR7 database and $d_{Car}$ the
distance given by Carollo et al. and derived from the \cite{Beers00}
sequences.  

The three branches, into which the sample stars are selected, can be identified
in the figure. Lower metallicity stars are brighter and bluer, but the
latter shift dominates. Thus the main sequence gets fainter
at the same colour. As the metallicity
spread of especially the main sequence in the top panel of
\figref{fig:HRrgi} partially obscures the underlying sequences, we
restricted the sample to $\feha < -1.9$
in the middle panel. For comparison we plot the adopted main sequence calibration
at $\feh \sim -2.14$ and the alpha enhanced Basti isochrones \citep[][]{Piet04,
  Piet06} for SDSS colours \citep[cf.][]{Marconi06} at $\feh = -2.14$ at ages $10$ and
$12.5$ Gyrs. The main sequence and the giant branch are
apparent and between them lies a strong sequence sloping down from $(0.2,
3.8)$ to about $(0.8, 4.8)$. The stars in this band are termed
turn-off stars by Carollo et al. as they comprise everything that has
gravities $3.5 < \llg < 4$.  As we can see from the bottom panel,
this artificial turn-off sequence comprises the majority of heavily
counter-rotating stars (green dots). The reader might ask why we do
not see a large number of stars as false identifications in the
brightest subgiant branch. As already discussed above the absolute magnitude difference to those
stars is, however, so large that by the distance limits
most stars sorted into the subgiant and giant branches will be dropped
from the sample, as they are deemed to be more than $4 \kpc$
away. This conclusion is verified by the subgiant branch getting
much more populated when the distance cut is removed (cf. Fig 2).

\begin{figure*}
\epsfig{file=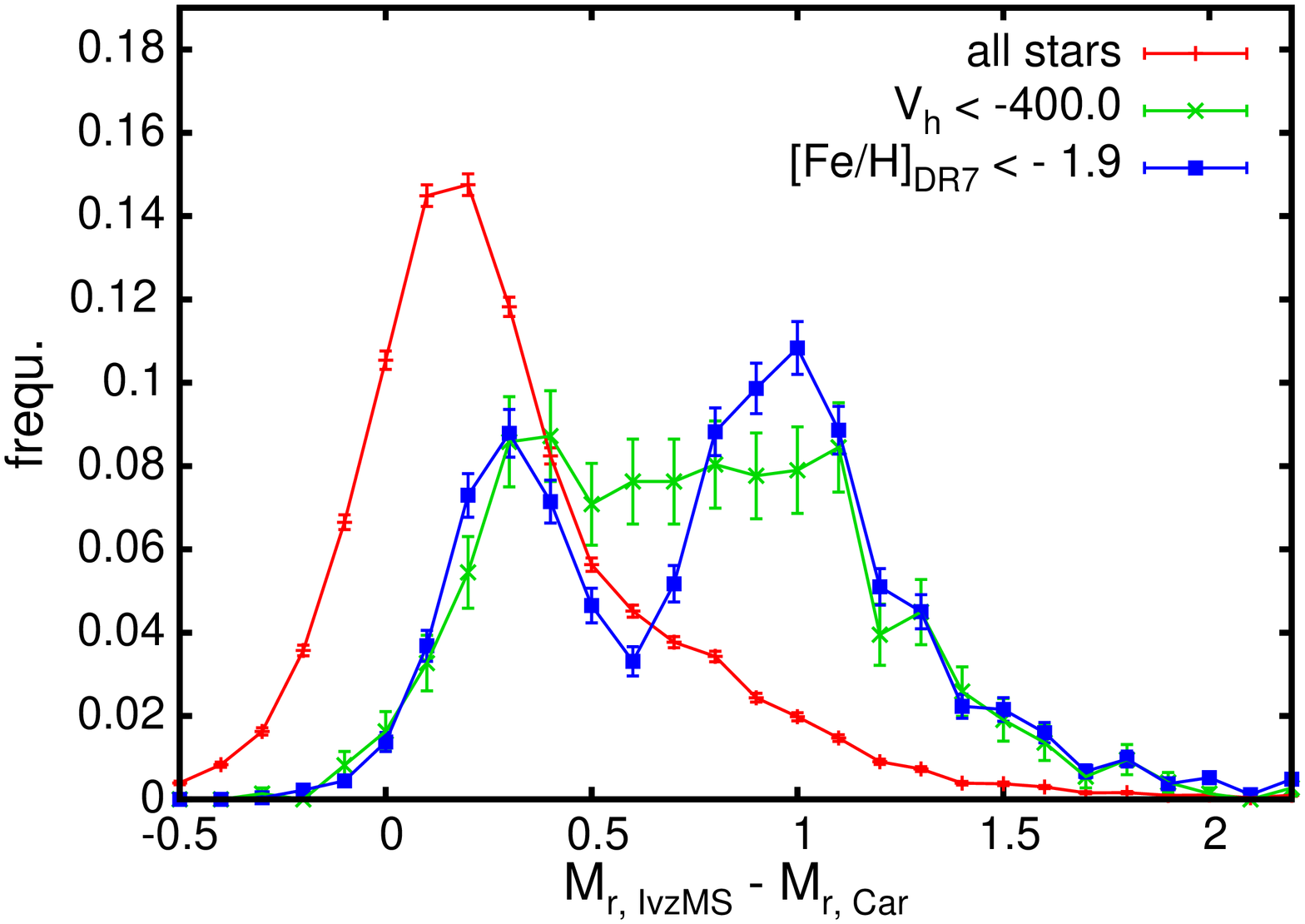,angle=-90,width=0.495\hsize}
\epsfig{file=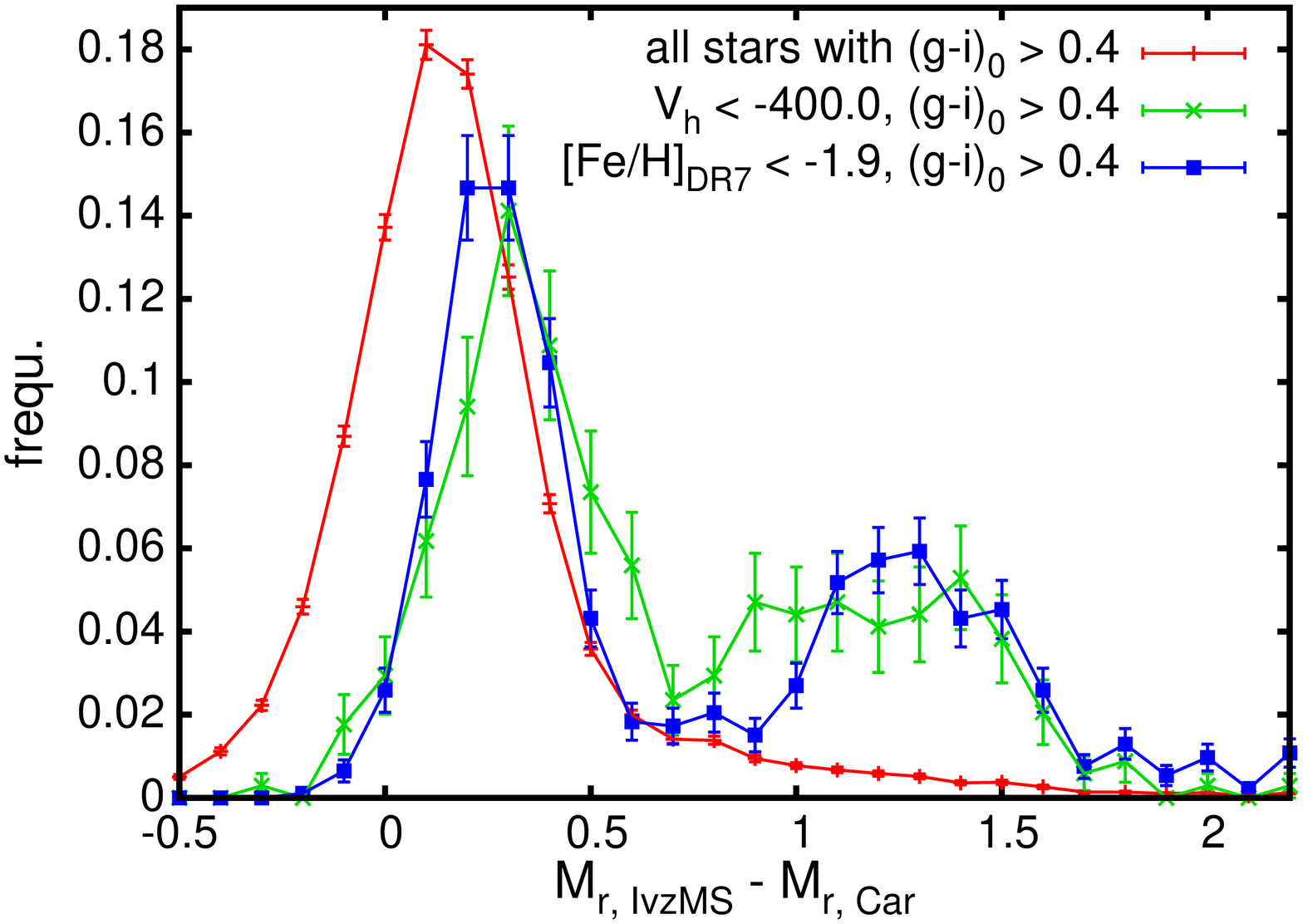,angle=-90,width=0.495\hsize}
\epsfig{file=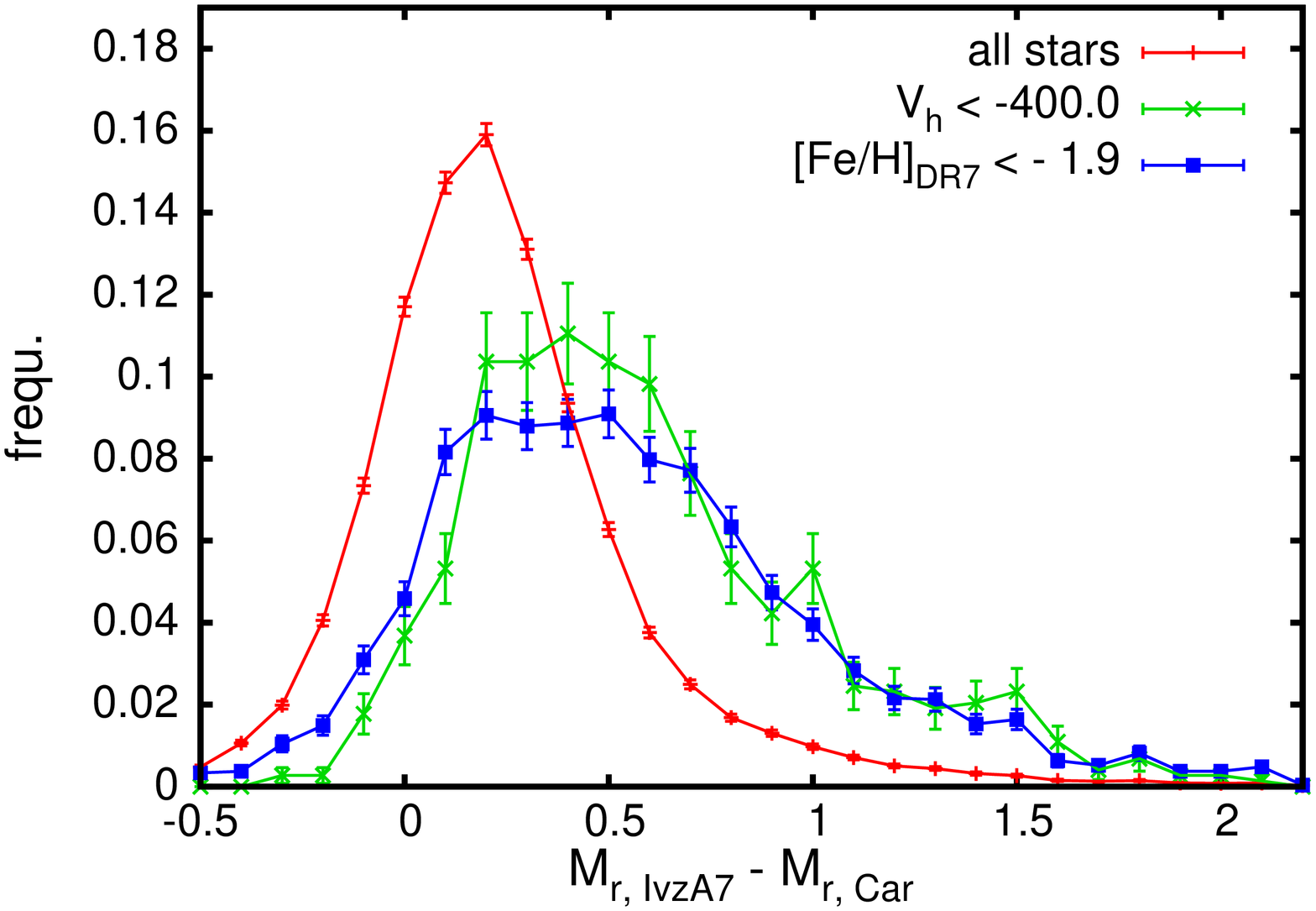,angle=-90,width=0.495\hsize}
\epsfig{file=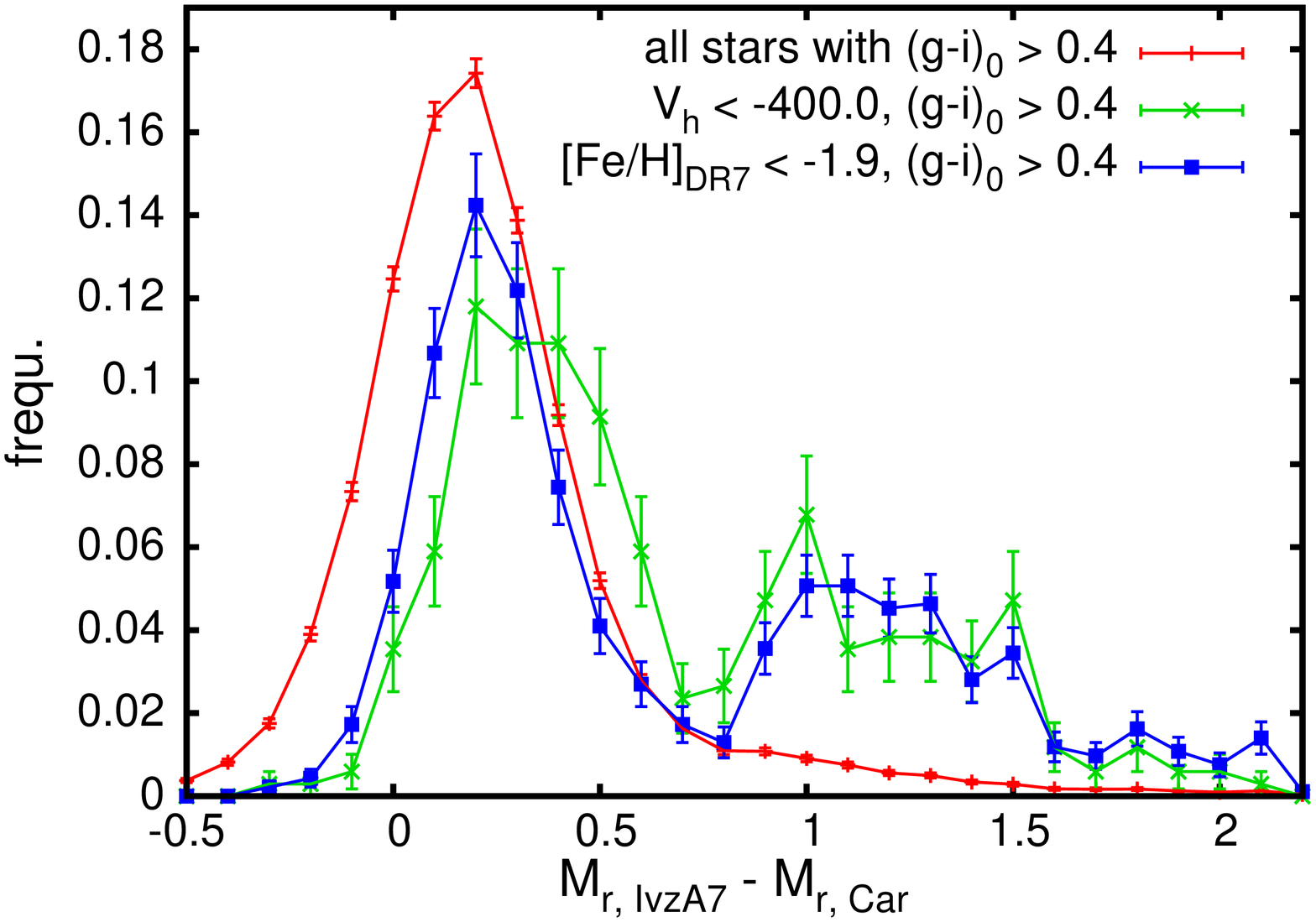,angle=-90,width=0.495\hsize}
\caption{
{The distribution of absolute magnitude differences between the Carollo et al. calibration and the other metallicity-dependent colour-magnitude calibrations. In the top row we show the results for the adopted main sequence calibration ("IvzMS"), while in the bottom row we compare to the age-dependent calibration by Ivezi{\'c} et al. (2008) ("IvzA7"). On the left we show the entire colour range, while we plot only redder objects with $(g-i)_0 > 0.4$ on the right. In the entire sample (red bars) the main sequence by Carollo et al. is brighter by $\sim 0.15 \mag$ than in the native SDSS calibrations. The offset gets larger for the counter-rotating (green crosses) and metal-poor (blue squares) stars; note that the Carollo et al. turn-off sequence is very prominent with an offset of about 1 magnitude.}
}\label{fig:rmagdiff}
\end{figure*}

In stellar evolution there exists no unique
turn-off branch. Instead there is just a region where stars leave the
main sequence and move up to the subgiant branch, and which depends on
metallicity and age of the population. Thus the artificial sequence we
face here can only be thought of as a compromise for stars in the broad region between main sequence
and subgiant branch that intends to describe the average luminosity of
these objects. In this perspective, the claim by Carollo et al. to
achieve $10-20\%$ accuracy in distances is ruled out in this
transition region as it intrinsically spans more than one magnitude
(at fixed metallicity and colour), i.e a distance uncertainty of more than $50 \%$.
Because this "turn-off" branch is constructed to be a compromise between
subgiants and main sequence, the gap between
the subgiant branch (moving up towards lower metallicity) and the main sequence (moving down, as
the colour effect dominates) widens with decreasing metallicity. This aggravates the
effects of misassignments on estimated distances for lower metallicity
objects.

The worst problem with the ``turn-off'' stars appears in comparison
with stellar models. Comparison of the adopted positions of metal-poor
stars in the colour-magnitude diagram with the isochrones at $10$ and $12.5$ Gyrs reveals that most objects in the artificial ``turn-off'' branch reside on the red side of the low metallicity turn-off region. Some of the selected 
``turn-off'' stars are even on the red side of what could be achieved
at solar metallicity. In fact the overwhelming majority of strongly
retrograde stars is actually claimed to be in a region where
according to our knowledge of astrophysics no star of reasonable ages
can reside.

\subsection{Colour transformation and main sequence comparison}

It should be mentioned that the distances used in
C07 and C10 were derived using the $B-V$
colour calibration and thus the colour transformation of \cite{Lee08a} had to be performed to apply them to SDSS
colours. Fortunately there are now  good isochrones
\citep[e.g. BASTI isochrones,][]{Piet04}, fiducials \citep[][]{An08}
and photometry in SDSS colours readily available, making such a
colour transformation to translate SDSS colours to the former $B-V$
calibration by \cite{Beers00} unnecessary. The colour transformation may explain some of the scatter and systematic shifts presented in \figref{fig:rmagdiff} that shows the comparisons of the absolute magnitudes from C10 to the native SDSS calibrations. The latter is depicted by red lines in \figref{fig:HRrgi} at $\feh = -2.14$ and $\feh = 0$. Both isochrones and the two main sequence approximations are fainter than the adopted
 C10 absolute magnitudes, especially at lower
 metallicities. This can be seen from the middle panel of
 \figref{fig:HRrgi} and from \figref{fig:rmagdiff}, which depicts the
 distribution of differences in absolute magnitudes of C10 towards those derived via the adopted main sequence calibration (top panels) and those derived via the \cite{Ivz08} age-dependent relation (bottom panels) for all stars (red), metal poor objects with $\feh < -1.9$ (blue) and for the "counter-rotating" stars with $V_{h}
 < -400 \kms$ (green crosses and errorbars). Errorbars depict the
 Poisson noise. On the right side of \figref{fig:rmagdiff} we show
 the same quantities, but exclusively for redder stars with $(g-i)_0 >
 0.4$ where the sequences of \cite{Beers00} have a larger
 separation, and which also excludes the expected turn-off region
 (cf. the isochrones in \figref{fig:HRrgi}) for the very metal-poor stars. Apart from the observational scatter, the peaks arising from the main sequence $\Delta_M \sim 0.3$ and from the "turn-off branch" around $\Delta_M \sim 1.0$ are washed out for the whole population, as the sequences shift with metallicity and the main sequence dominates. However, the generally higher intrinsic brightness under the Carollo et al. assumptions is clearly seen in both distance descriptions. On the blue end the age-dependent formulation by \cite{Ivz08} is a bit brighter than the adopted main sequence calibration and makes the peak of the "turn-off sequence" in the metal poor and counterrotating subsamples of C10 merge with the main sequence at blue colours, yet the large offset remains.

For stars that are claimed by C10 to have heliocentric velocities $V_{h}
  < -400 \kms$ and thus make up the bulk of what they identify as a counter-rotating halo  the two peaks of main sequence and turn-off stars are
  separated, as the metallicities of these stars are more narrowly
  distributed . For comparison we show the result of selecting only stars
  that have $\feha < -1.9$ (blue squares in \figref{fig:rmagdiff}). These distributions look very similar apart from the counter-rotating stars being more inclined to the bright side. Also regard the increased importance of the turn-off band for those populations.

In summary there are two main drivers of overestimated distances in
Carollo et al. compared to the native SDSS distance calibrations: The main sequence of
C10 for metal-poor stars is brighter than in the other calibrations and the number of stars in the turn-off and subgiant/giant
branches is increased by a large factor for metal-poor stars; most of the designated ``turn-off'' stars by C10 reside in positions in the colour-magnitude plane that would require unreasonable ages.  
We thus conjecture that the ``counter-rotating'' halo as presented by C10 is due to distance
uncertainties and the selection of stars into unphysical stellar
branches an artifact.

\begin{figure}
\epsfig{file=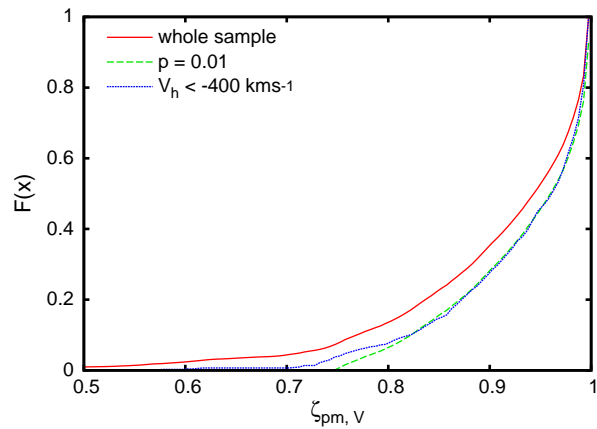,angle=-90,width=\hsize}
\caption{The cumulative distribution of stars on the part of $V_h$ that is covered by proper motions and not radial velocity measurements. The stars with $V_{h} < -400 \kms$ are drawn with a purple line while the overall sample distribution is drawn with a red line. Beneath is the $0.01$ significance level of a Kolmogorov-Smirnov test.}\label{fig:advgeo}
\end{figure}

\section{Signatures of misselection}\label{sec:sign}
\subsection{Geometric distribution anomaly}

Even if the biases in distance assignments were not obvious there are
still ways to detect them. As erroneous distance estimates mostly
act on the part of motion perpendicular to the line of sight,
populations with distance errors will be
preferentially found in regions of the sky perpendicular to the
velocity component in question, or in other words, if stars end up in
one component by misselection this should should show up as a bias in
sample geometry. Such a bias will not be aligned directly with the
Galactic coordinates, but with the part of the biased velocity component
covered by proper motion. This is obvious as the largest errors should
happen where the uncertainty in the motion is largest, i.e. where the
proper motion and distance estimates have the largest impact.
For these statistics we make use of the squared angle terms connecting
the proper motion to the azimuthal velocity (from now called the "proper motion partition"). We take from eq.\eqref{eq:vvh} in the Appendix the angle terms that connect $V_h$ to the proper motions ${\dot l}$ and ${\dot b}$, square them and add them together:
\begin{equation} 
\zeta_{pm,V} = \sin^2(l)\sin^2(b) + \cos^2(l) .
\end{equation}
 \figref{fig:advgeo} shows the cumulative distribution of stars over
 the proper motion partition for $V_h$ (red line). The C10 sample is (as the entire SEGUE survey)
 concentrated towards positions where the azimuthal velocity is mostly
 covered by proper motions as it is mostly oriented towards high
 Galactic latitudes. Yet the counter-rotating subsample (purple line)
 is even more concentrated towards the high proper motion
 contributions. The average value for $\zeta_{pm, V}$ rises from $0.909$
 for all stars to $0.933$ for the subsample with $V_{h} <
 -400.0$. In other words, the fraction covered by robust radial
 velocities drops from $0.091$ to $0.067$. In a Kolmogorov-Smirnov test the probability for equality of the two distributions is well below the $1 \%$ level and thus equality is strongly rejected. Again
 the high values for $\zeta_{pm,V}$ show how vulnerable the sample azimuthal velocities are to any distance errors.

This is of course not a proof of the bias, but a strong indication. Against this argumentation one could raise the objection that stars at low metallicities are located more polewards, as the halo to disc ratio in the sample rises (higher altitudes are reached) and so metal-poor / halo stars show a different spatial distribution. Indeed for lower metallicities the sample distribution shifts polewards reducing the difference, which remains present in all cases, but can become insignificant due to the shrinking sample numbers. 

\begin{figure}
\epsfig{file=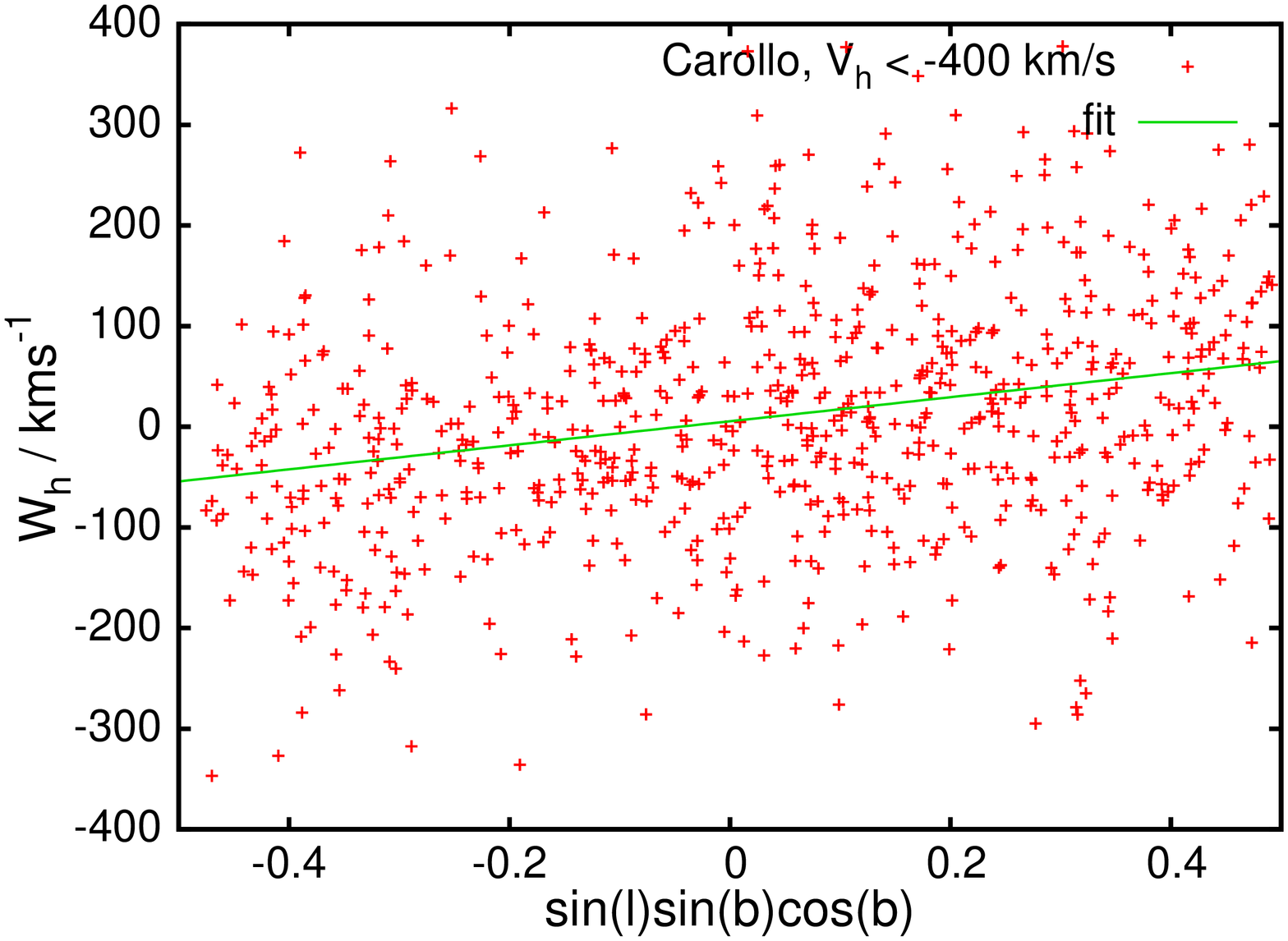,angle=-90,width=\hsize}
\epsfig{file=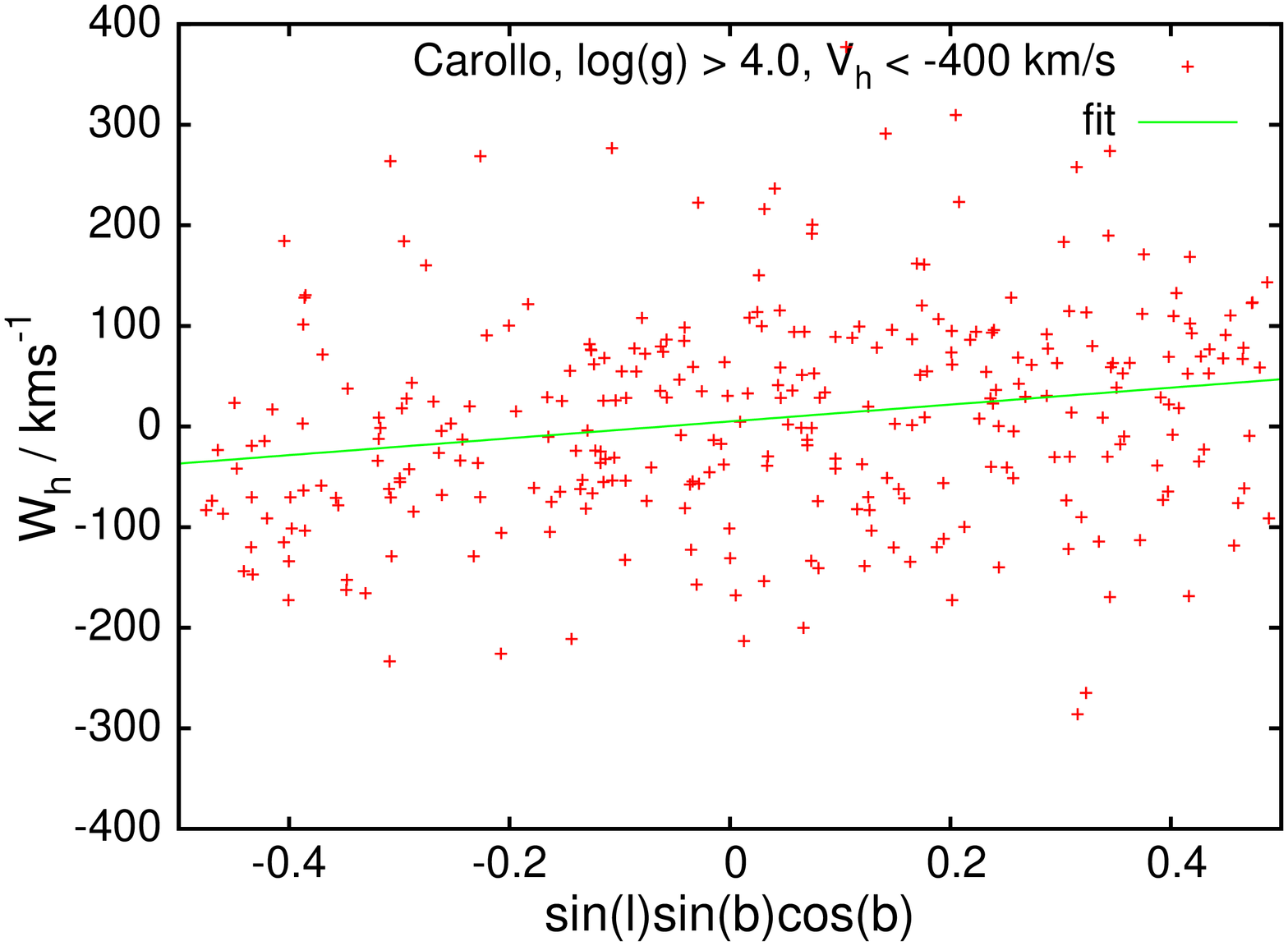,angle=-90,width=\hsize}
\caption{The counter-rotating stars display an intense trend in average
$W_{h}$ velocity over the angle combination that describes the connection of the azimuthal and vertical velocity components via distance errors.}\label{fig:WvsV}
\end{figure}
\subsection{Linear error analysis and velocity crossterms}\label{sec:crossterms}

A robust approach to prove and quantify the distance errors is presented in
\figref{fig:WvsV}, which shows the $W_{h}$ velocities
against $\sin(l)\sin(b)\cos(b)$ for all stars that should be part of
the counter-rotating halo, i.e. have $V_h < -400 \kms$. The strong
uptrend cannot be any stream, which would show up as a narrower band. It appears that the entire halo
experiences a lift in $W$ velocity, which has its origin in the distance
errors of the C10 analysis. By wrongly assessing distances the relative motion of stars in
the azimuthal direction mixes over the proper motions into the
vertical component depending on this angle term. To illustrate this
effect think of a star that is seen at $b = 45^{o}$, $l = 90^{o}$ and
has $U_h = 0$, $V_h = -500 \kms$, $W_h = 0$. Due to the geometry of
the setup the star will have a significant proper motion in $b$. With a wrong distance estimate the relative impact of line-of-sight
velocity and proper motion changes and the star is
assigned a non-zero $W$ velocity. A more detailed description is found in Sch\"onrich et al. (in prep.). 
The fitting line has a slope of $119.5 \pm 16.5 \kms$, i.e. the trend is significant at a level of more than $7 \sigma$. 
We can thus state that a significant distance error that is expected
from the discussions above is here quantitatively demonstrated. Since
the alleged counterrotating halo stars in this subsample (with a cut at $V_h < -400 \kms$) have on average
$\overline{V_{h}} \sim -470\kms$, we arrive at a distance bias of
roughly $40 \%$ for the counter-rotating stars, which translates into
a magnitude error of around $0.7 \mag$. It should be mentioned that
errors might even be higher: Selecting more strongly retrograde stars
the trend estimate gets even higher.

Restricting the sample of stars with $V_h < -400 \kms$ further to dwarfs ($\llg > 4.0$) and thus removing the largest identified source of distance errors, the numbers of stars in the
sample drop by the selection dramatically from $735$ stars to $299$. The removal of the low gravity stars also gives fewer
outliers in the velocity distribution. Only one object has $|W| > 400
\kms$. The lower panel of \figref{fig:WvsV} shows the remaining stars
together with an equivalent linear fit. Again the trend is highly
significant, though now more moderate with $83.8 \pm 22.5 \kms$. The
$\sigma_W$ in the sample is already down from $129 \kms$ to $110 \kms$. The more
moderate trend corresponds to an error of about $25 \%$. It also fits
well into the picture that the distance overestimate for the full
sample is larger than for the dwarfs, when we remove the spurious
"turn-off" branch. This bias has the effect that the
distance overestimates do not only increase the $W$ velocity directly, but also give rise to an additional contribution to the $W$ velocity dispersion by turning a part of the large heliocentric azimuthal motion of halo stars into a fake vertical term. We will call this behaviour ``velocity crossovers''.
This effect is even more important in the original analysis
by C07 and C10, as they restrict their sample in galactocentric
radius, thus increasing the weight around $l = 90^{\circ}$ and $l =
270^{\circ}$, where $\sin(l)$ is largest.

\begin{figure}
\epsfig{file=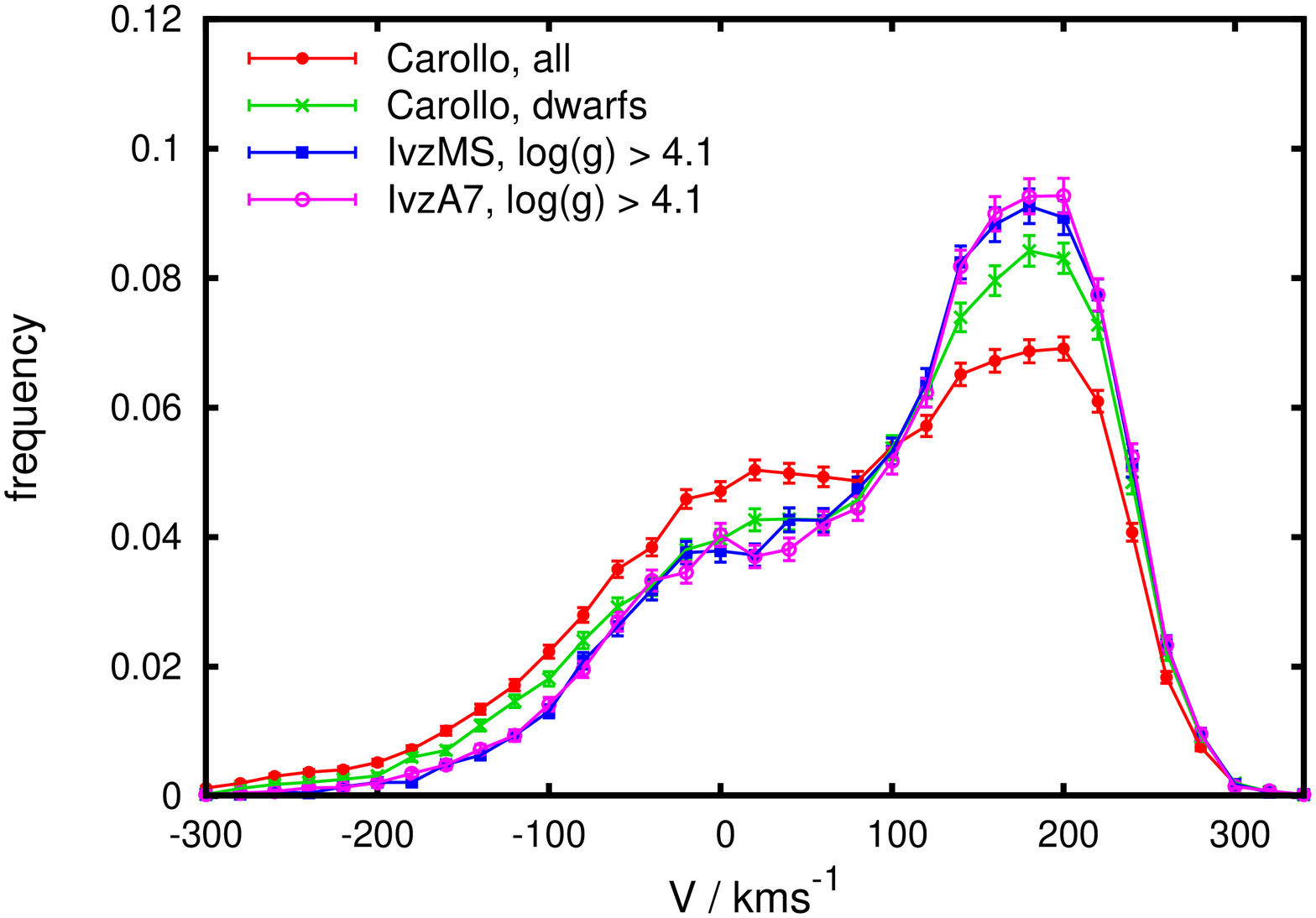,angle=-90,width=\hsize}
\epsfig{file=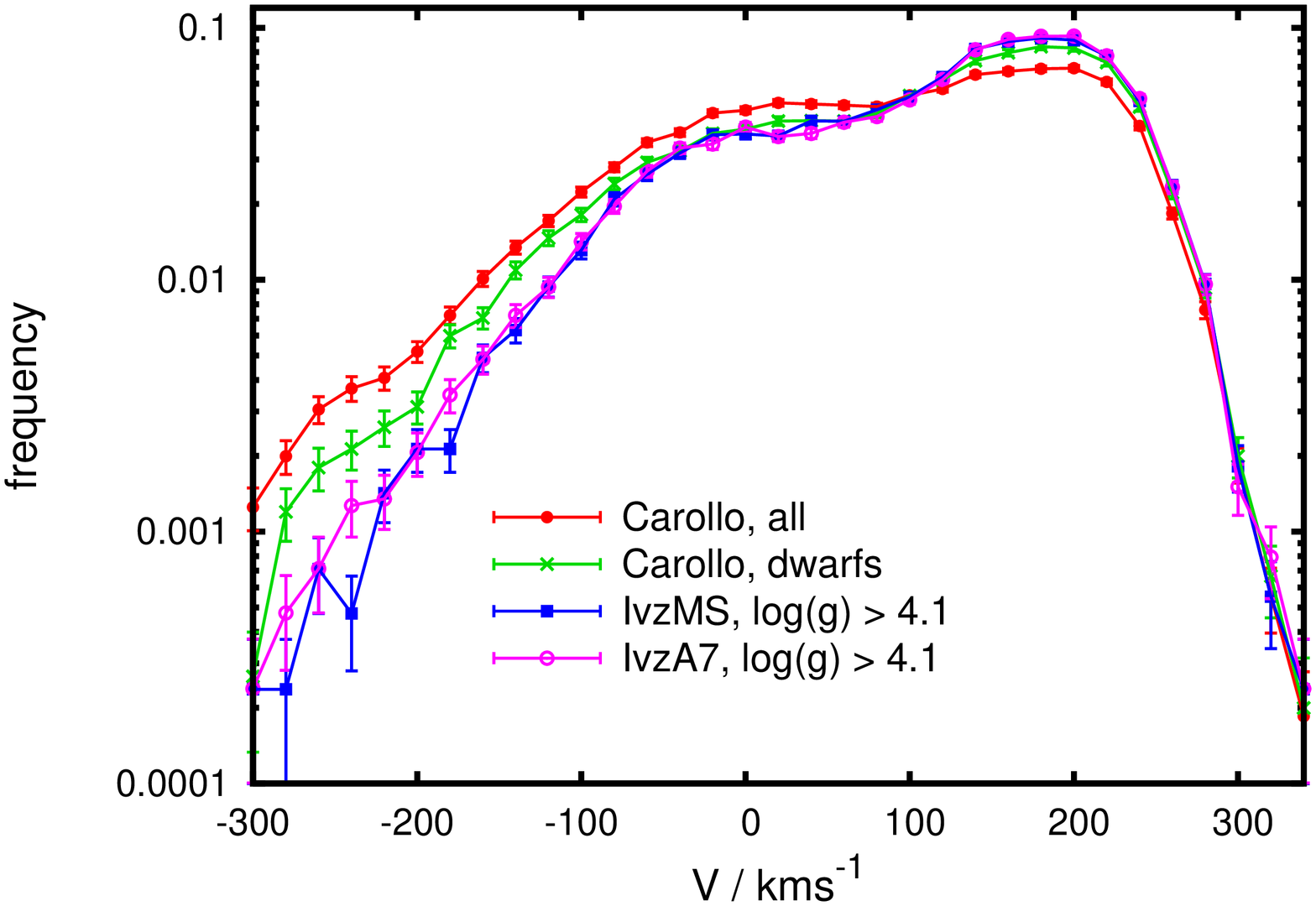,angle=-90,width=\hsize}
\caption{Velocity distributions for stars in the calibration sample. To make them comparable at the different sample sizes, all distributions were normalized to unity (i.e. divided by the total number of stars in each subsample). Error bars show again Poisson errors.
}\label{fig:velos}
\end{figure}

\section{Velocity distribution}\label{sec:vel}

How do the described biases and different distance derivations affect the velocity distribution and what might remain of the alleged counter-rotating component halo when we reduce the distance biases that spuriously inflate it?
\figref{fig:velos} shows the velocity distributions under different
cuts and distance approximation both on a linear scale and on a
logarithmic scale that reveals the wings of the distributions. Error
bars give the Poisson noise, but neglect any other sources of
error. The velocity distributions of the halo and the Galactic disc
can be made out in both plots. The long tail of the counter-rotating
halo is clearly visible for the C10 sample (red circles). As already noted above,
a restriction to dwarf stars (green crosses) diminishes the counter-rotating tail (around $V < -170 \kms$, which corresponds to $V_h < -400 \kms$). The blue filled squares show what we obtain with the adopted main sequence calibration, and the purple empty circles depict the age-dependent distance calibration by \cite{Ivz08}.
With both native SDSS distance calibrations the sample displays a clean downtrend to the low velocity side corresponding to a $V$ velocity dispersion of around $70-80 \kms$.

One might argue that the two native SDSS calibrations lead to some
contamination by giants that are treated as dwarfs and thus provoke the
opposite effect of underestimating distances. To reduce the contamination by evolved stars we use a gravity cut of $\llg > 4.1$ in the following discussions, which mostly helps to reduce the density saddle between the disc and halo component, where halo stars with severe distance underestimates tend to assemble.  However, the left tail of the halo distribution does not change significantly on a tightening or loosening of the gravity cut. In the following discussions we will always use the slightly tightened condition $\llg > 4.1$, but we checked that all of our conclusions are valid regardless of the specific choice of the gravity cut.
The absence of a significant excessive tail for dwarf stars can in our
view only be explained by the fact that strongly counter-rotating halo
dwarfs are at the best an extremely rare population. It is highly
implausible that the counter-rotating halo consists exclusively of
subgiants and giants.

\begin{figure}
\epsfig{file=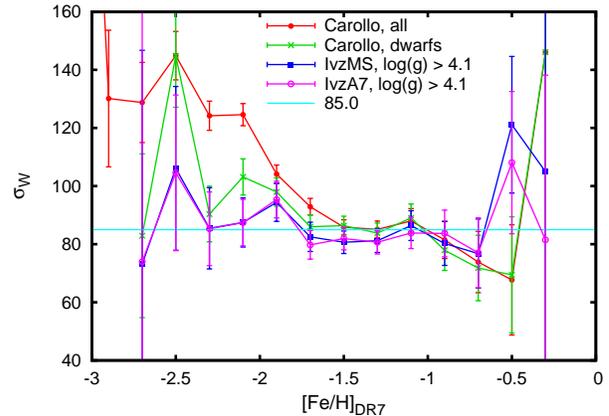,angle=-90,width=\hsize}
\caption{Vertical velocity dispersions against metallicity for different halo subsamples and distance calibrations. In all samples we cut away all prograde stars to eliminate most of the disc contamination. While the Carollo et al. full sample (red line) shows a vigorous uptrend towards the lowest metallicities, this trend almost vanishes in their dwarf star subsample (green line), while using the two native SDSS calibrations no significant trend is detectable.
}\label{fig:vertkin}
\end{figure}

\subsection{Kinematics versus metallicities}

The reason why Carollo et al. (2007, 2010) get a transition between their inner and
outer halo from a local sample is the increasing scaleheight of their
populations with lower metallicity, which is linked to the increasing
velocity dispersion perpendicular to the plane. According to our above
discussion of kinematic fingerprints of distance errors, at least the
increase of vertical dispersion by velocity crossovers (Section \ref{sec:crossterms}) due to distance overestimates should disappear when cutting away the unphysical turn-off stars and further when switching from C10 dwarf distances to the native SDSS calibrations. This is indeed observed in \figref{fig:vertkin}: It shows the values of vertical velocity dispersion against metallicity for different subsamples together with Poisson errors. The full C10 sample (red circles) harbours a prominent uptrend towards lower metallicity, which is at odds with the earlier result from \cite{Chiba00}. This uptrend almost vanishes in their dwarf subsample (green crosses). Both when using the adopted main sequence calibration (blue filled squares) and when using the \cite{Ivz08} age-dependent distance estimates (purple empty squares), we see no significant trend with metallicity any more; this finding is robust against changes in the gravity cut. To guide the eye we plot a horizontal line at $85 \kms$. There is a suggestion of a subtle increase below $\feha < -1.8$, yet numbers are too small to allow for a judgement. A source of uncertainty is the degree of possible giant contamination. It likely rises towards the metal-poor side, which may result in a reduction of measured dispersions. On the other hand the radial velocity support for the vertical velocity component is very high due to the polewards orientation of the sample. So the decrease effected by distance underestimates is relatively weak. Underestimating distances takes a further moderated effect on vertical dispersions: While a distance overestimate both increases the measured vertical dispersion by direct overestimate of the proper motion part, and a lift via the velocity cross-overs, on distance underestimates the effects have opposite signs, thus to some part balancing each other.

\begin{figure*}
\begin{tabular}{c|c|c}
{\large \bf $\quad \quad$ kin. energy} & {\large \bf $\quad \quad$ vertical kinetic energy} & {\large \bf $\quad \quad$ vert. kin. energy for retrogr. stars}\cr
\epsfig{file=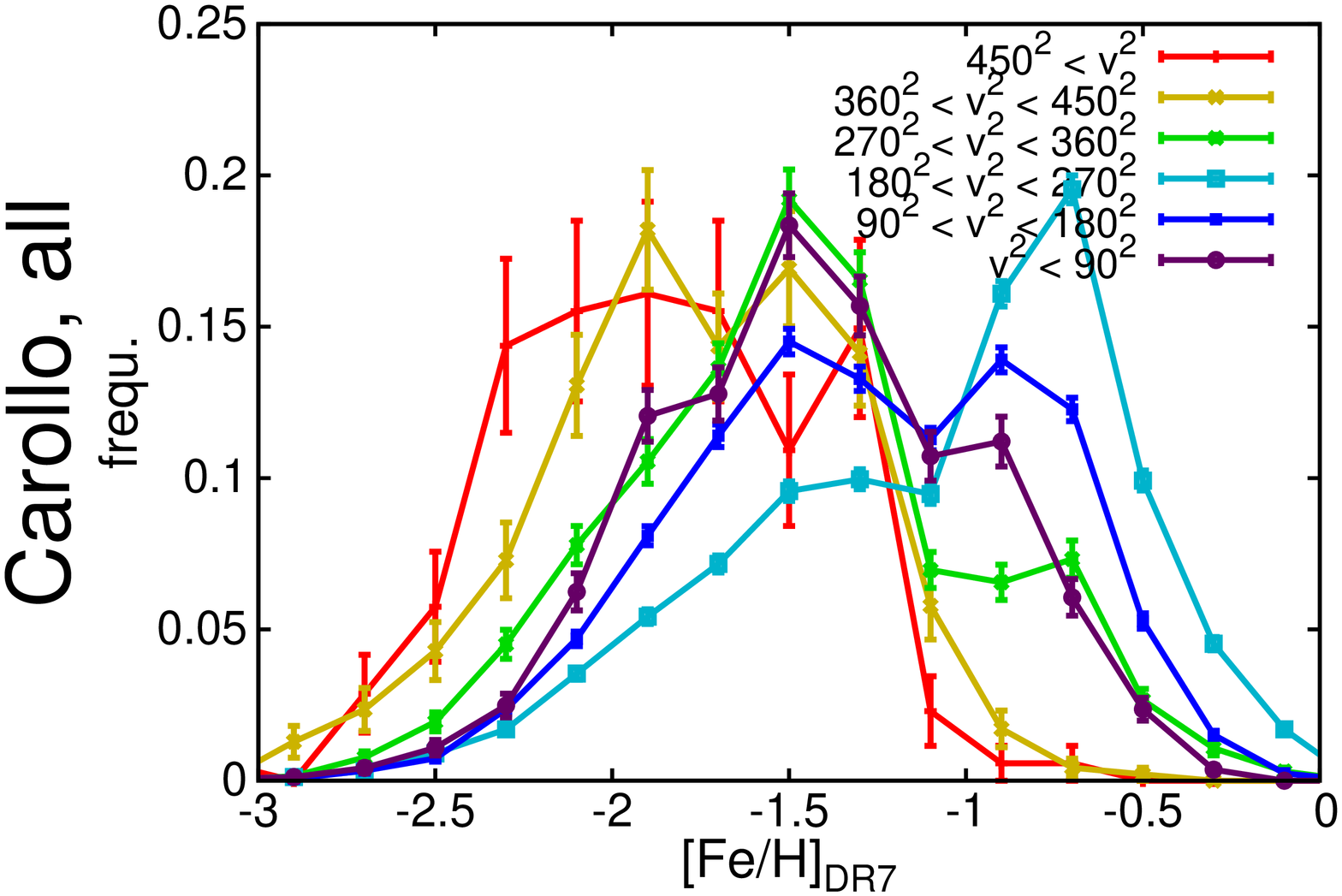,angle=-90,trim= 5mm 4mm 10mm 6mm,clip, width=0.35\hsize}&
\epsfig{file=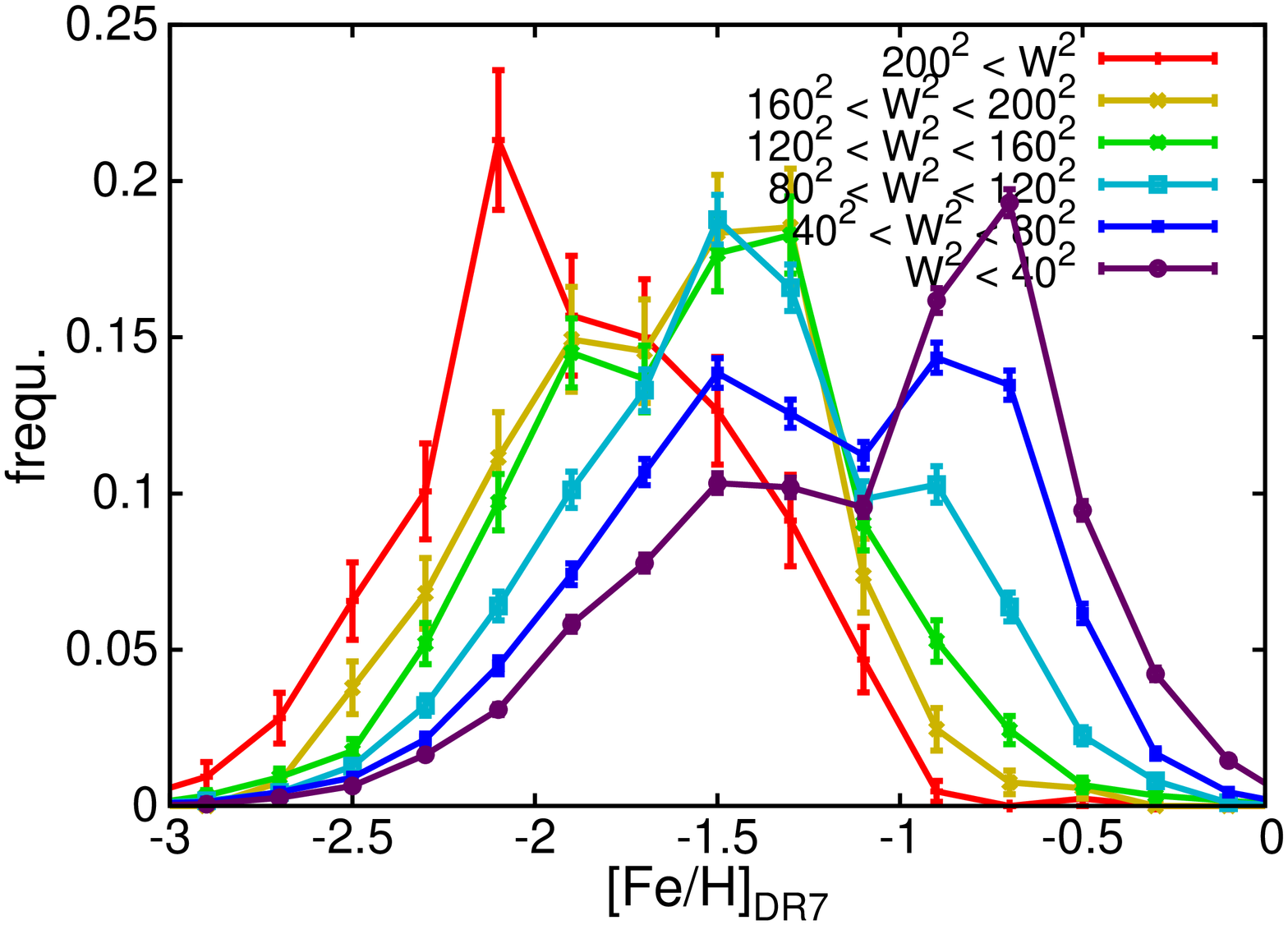,angle=-90,trim= 5mm 22mm 10mm 6mm,clip,width=0.325\hsize}&
\epsfig{file=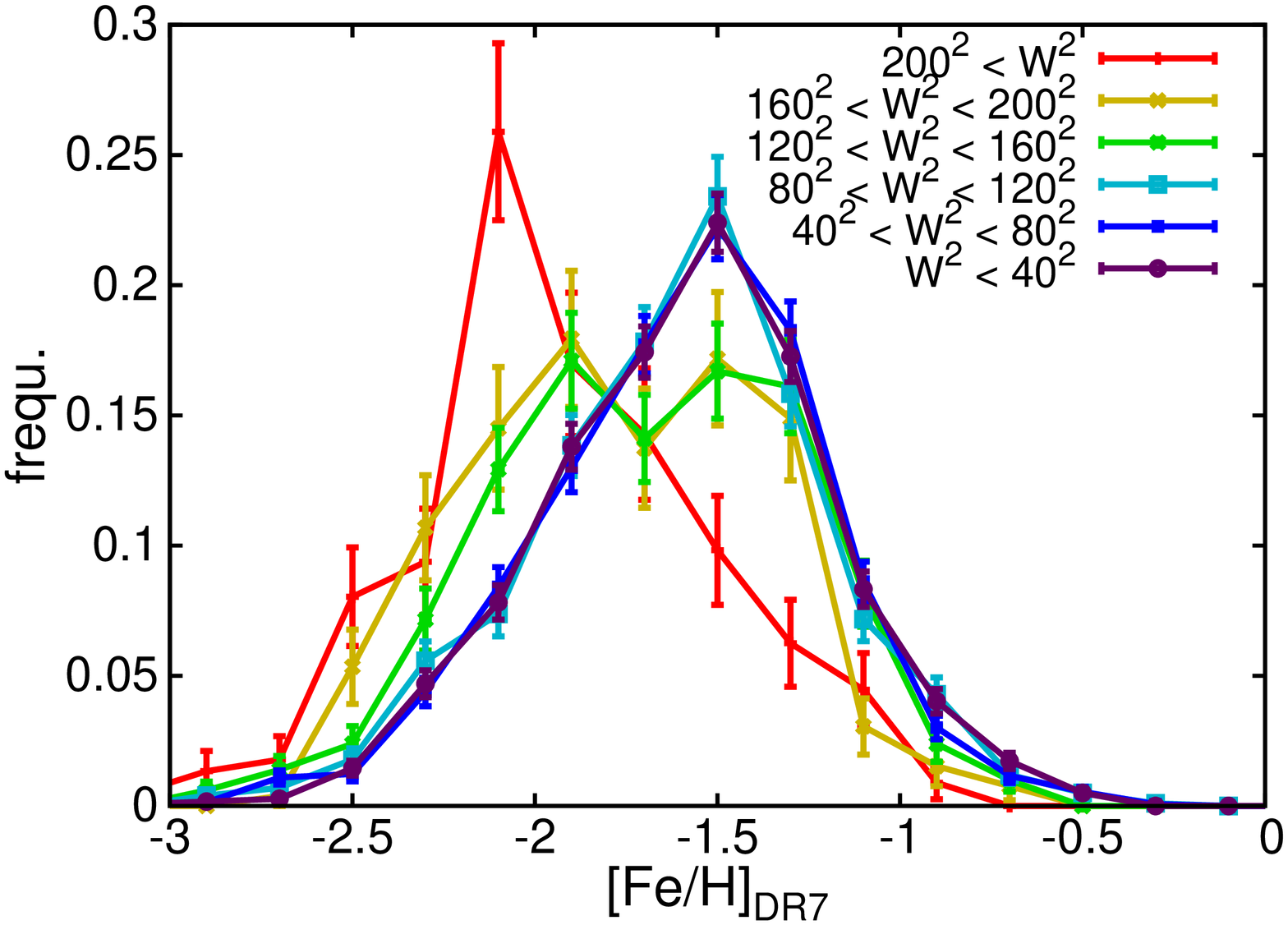,angle=-90,trim= 5mm 22mm 10mm 6mm,clip,width=0.325\hsize}\cr
\epsfig{file=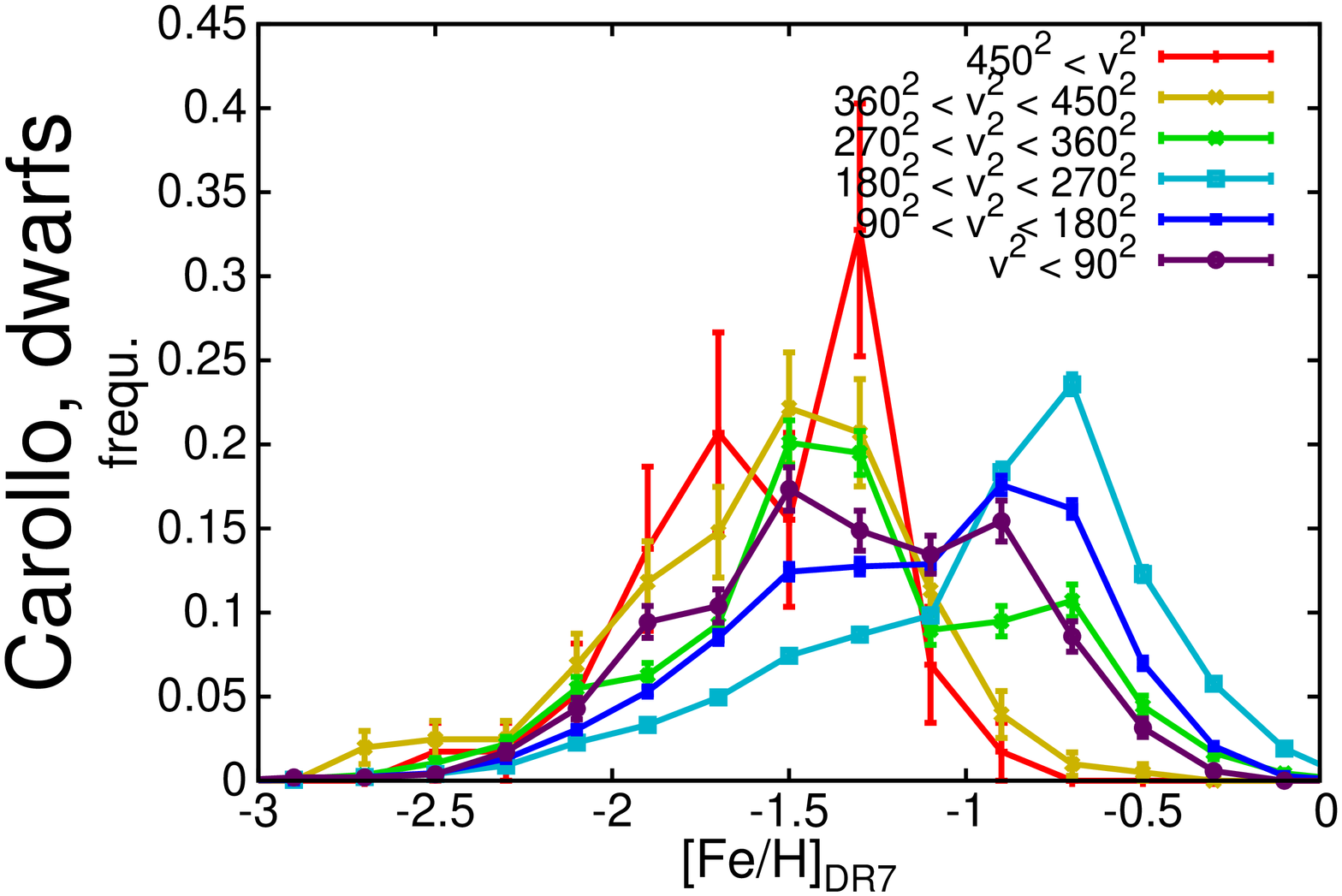,angle=-90,trim= 5mm 4mm 10mm 6mm,clip, width=0.35\hsize}&
\epsfig{file=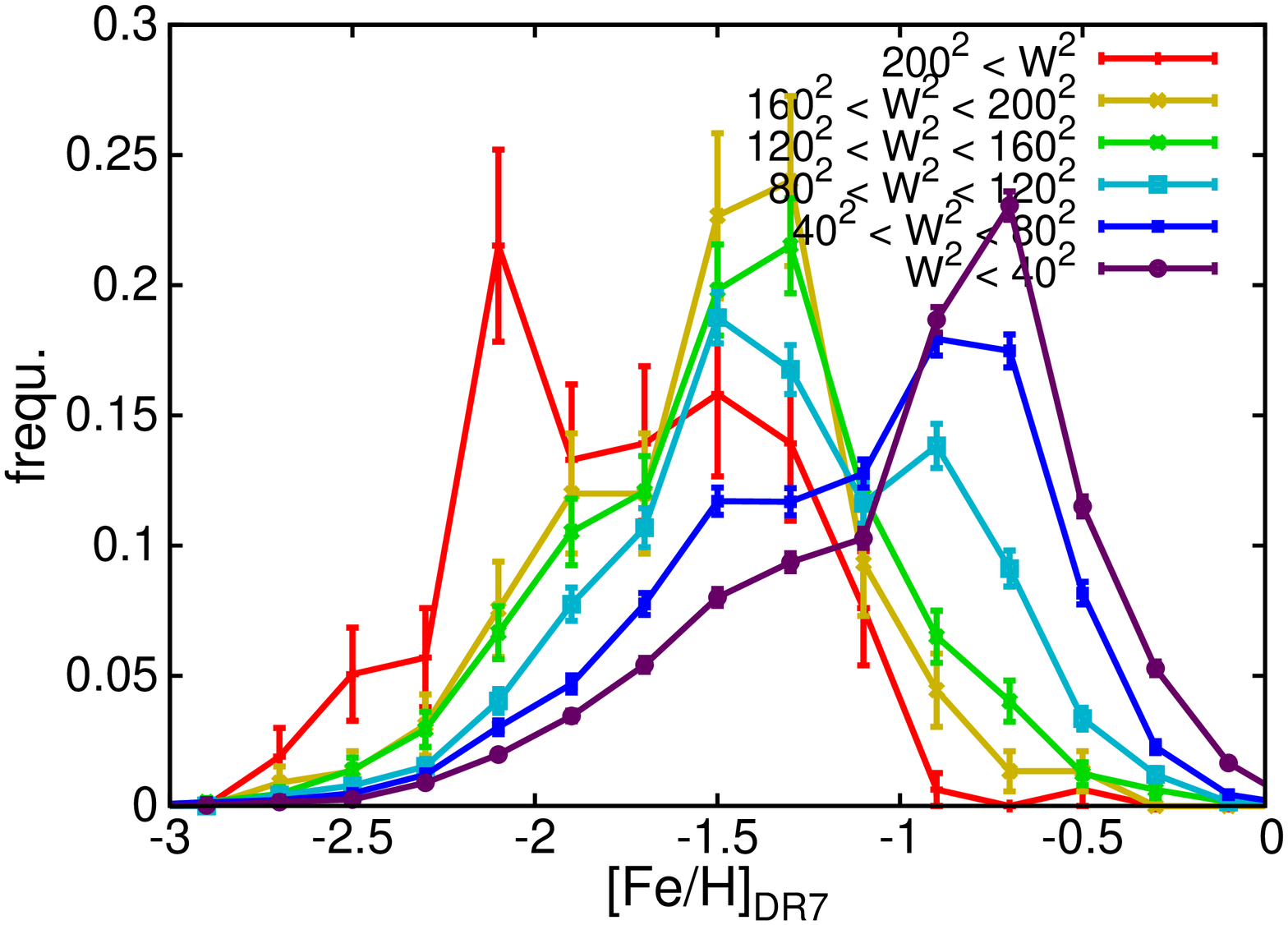,angle=-90,trim= 5mm 22mm 10mm 6mm,clip,width=0.325\hsize}&
\epsfig{file=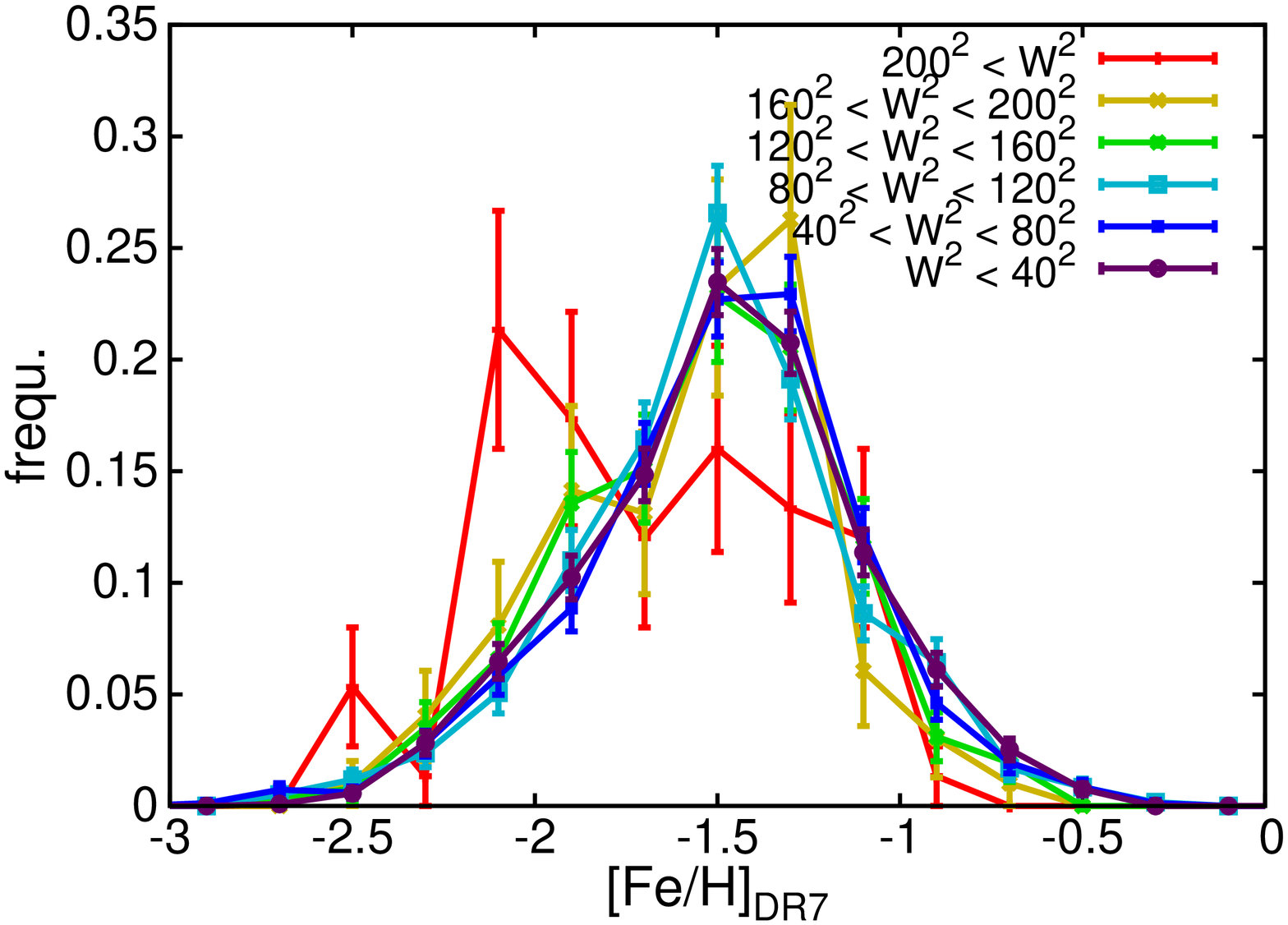,angle=-90,trim= 5mm 22mm 10mm 6mm,clip,width=0.325\hsize}\cr
\epsfig{file=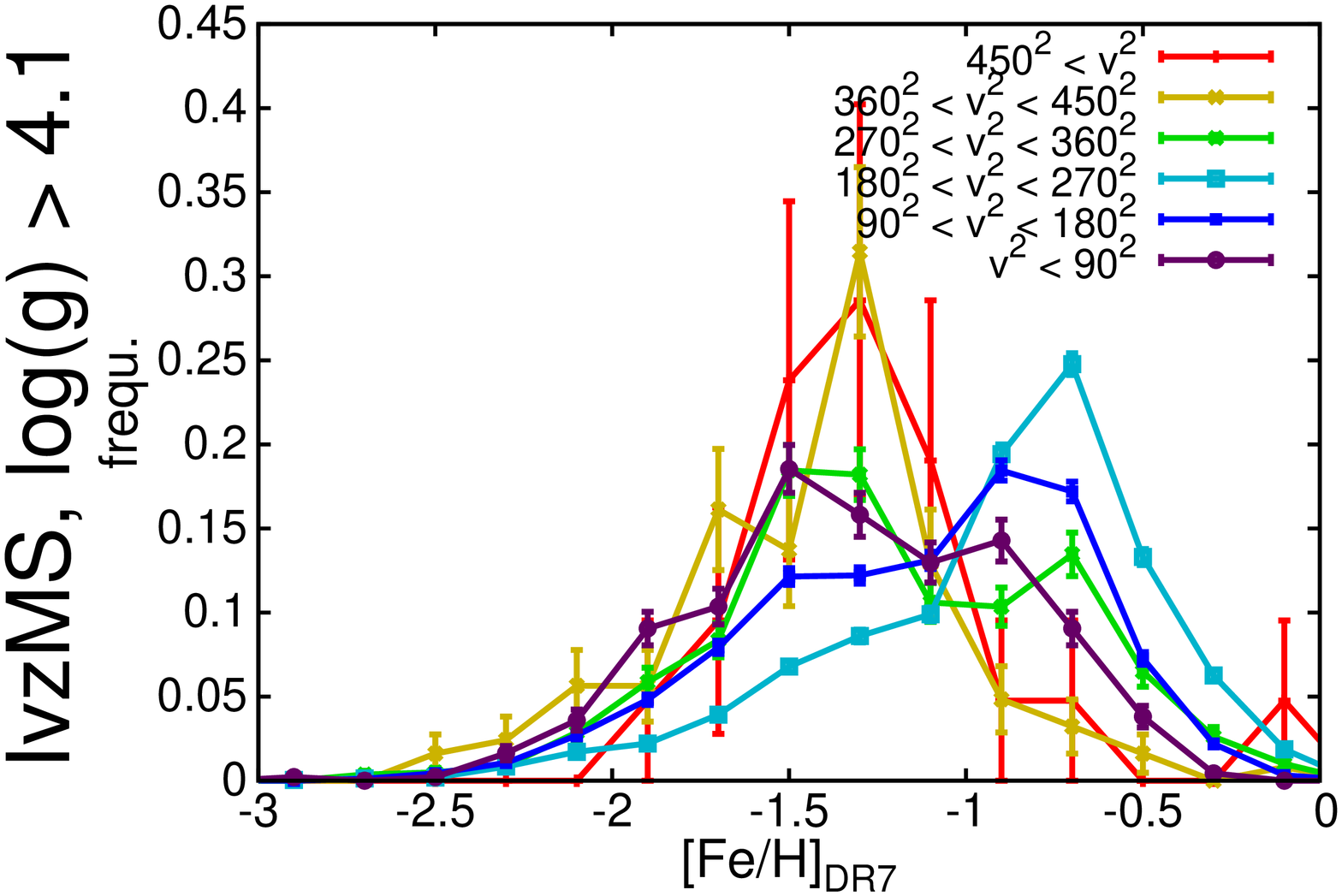,angle=-90,trim= 5mm 4mm 10mm 6mm,clip,width=0.35\hsize}&
\epsfig{file=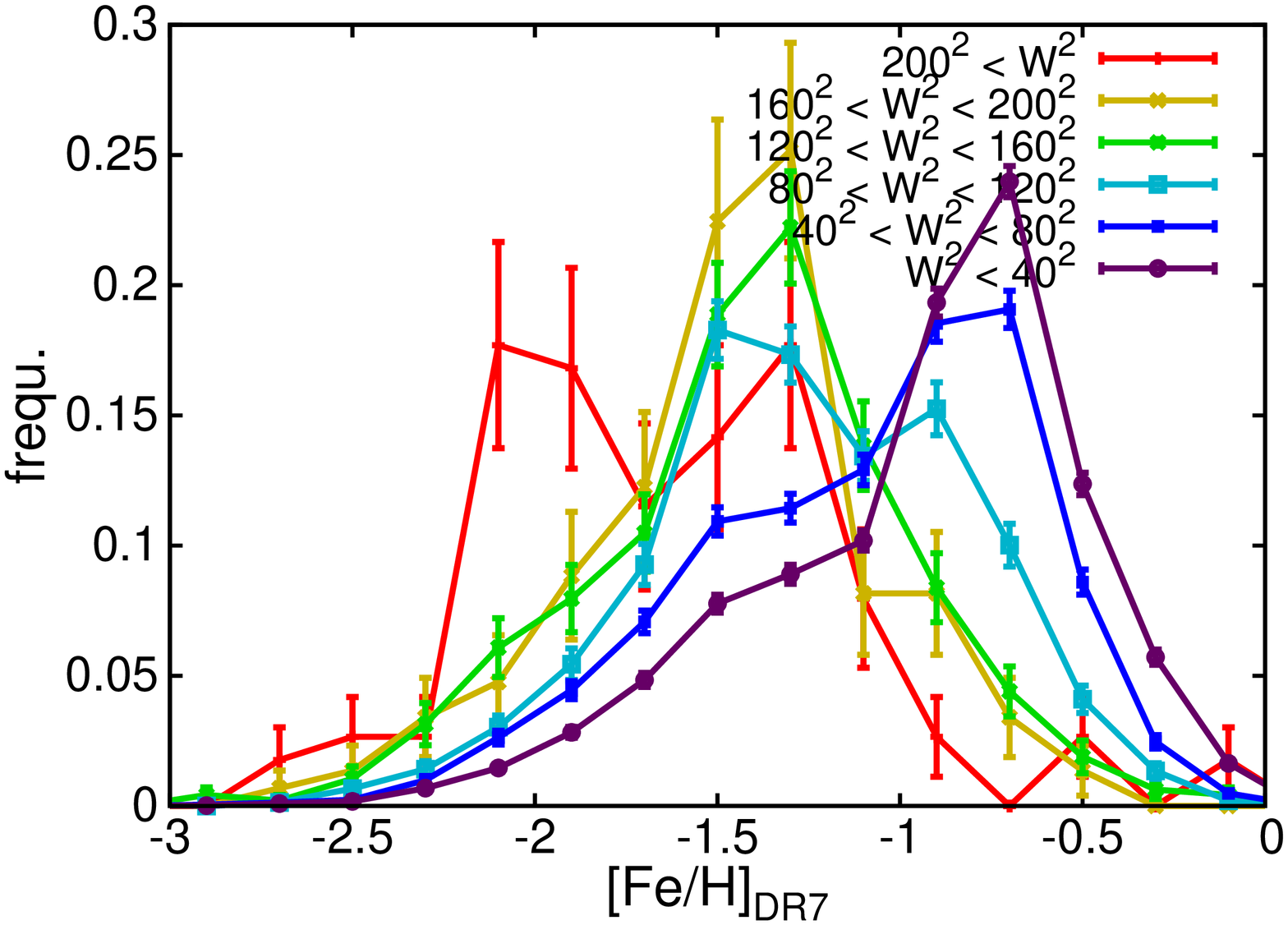,angle=-90,trim= 5mm 22mm 10mm 6mm,clip, width=0.325\hsize}&
\epsfig{file=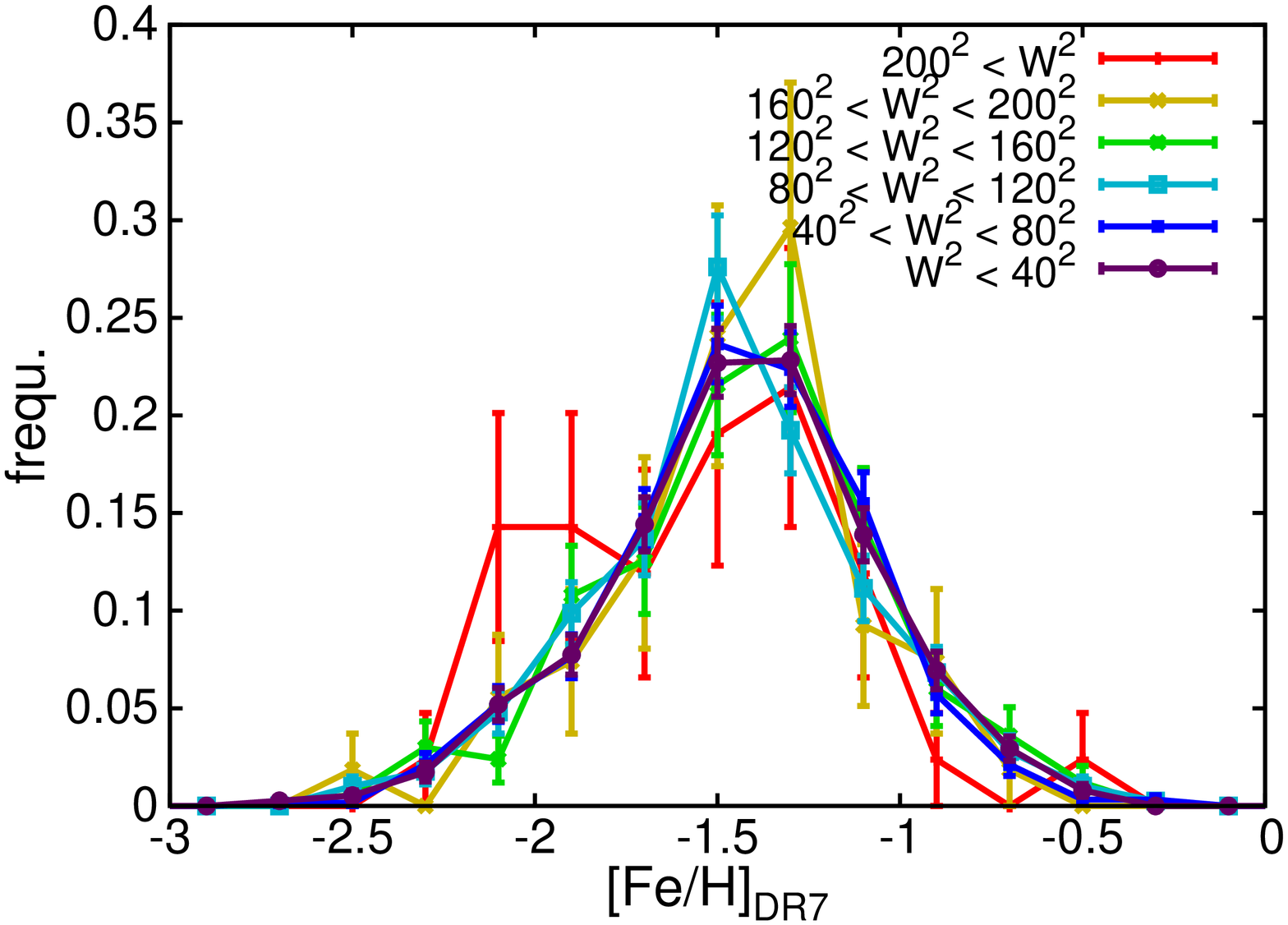,angle=-90,trim= 5mm 22mm 10mm 6mm,clip,width=0.325\hsize}\cr
\epsfig{file=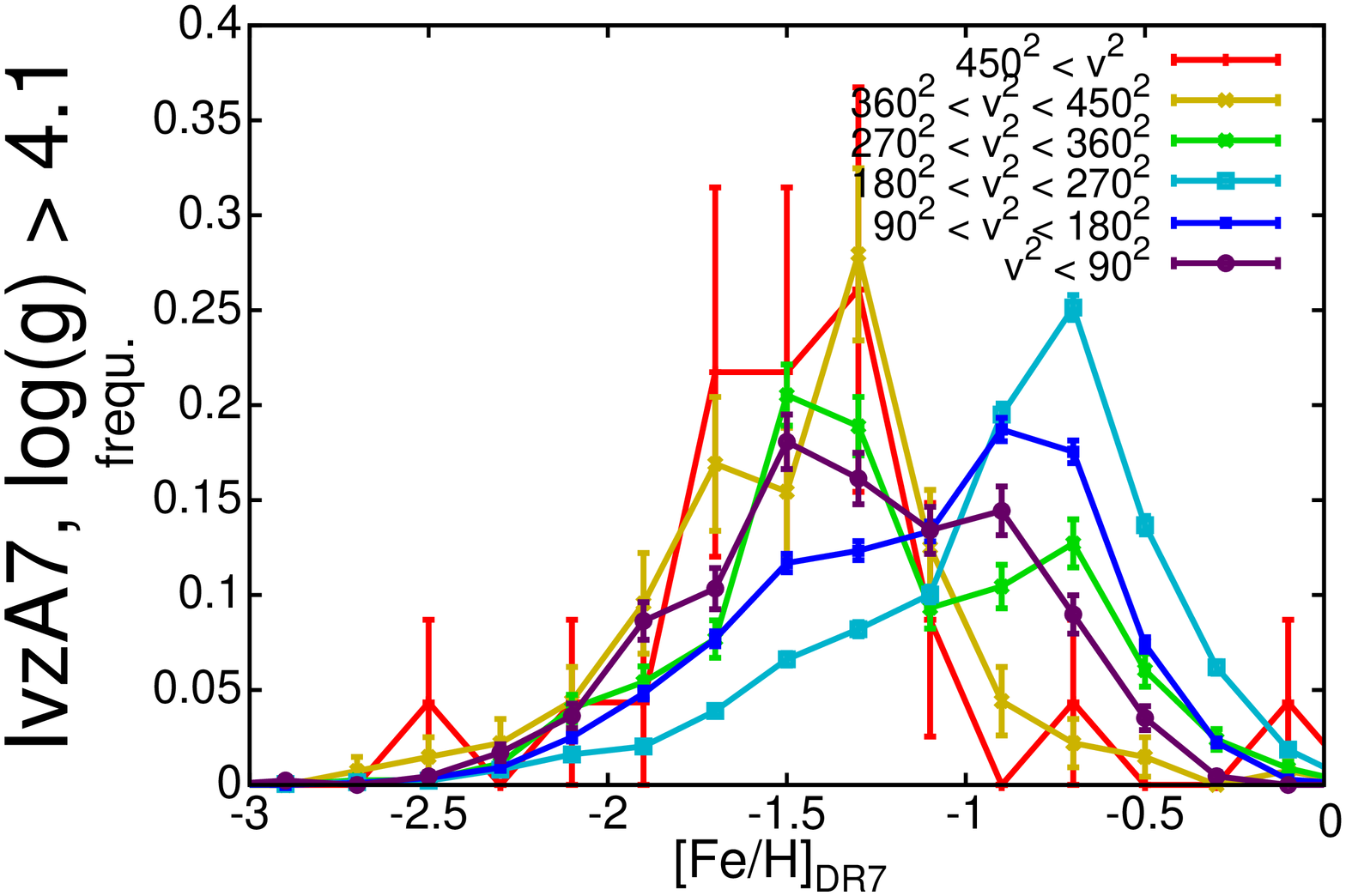,angle=-90,trim= 5mm 4mm 10mm 5mm,clip, width=0.35\hsize}&
\epsfig{file=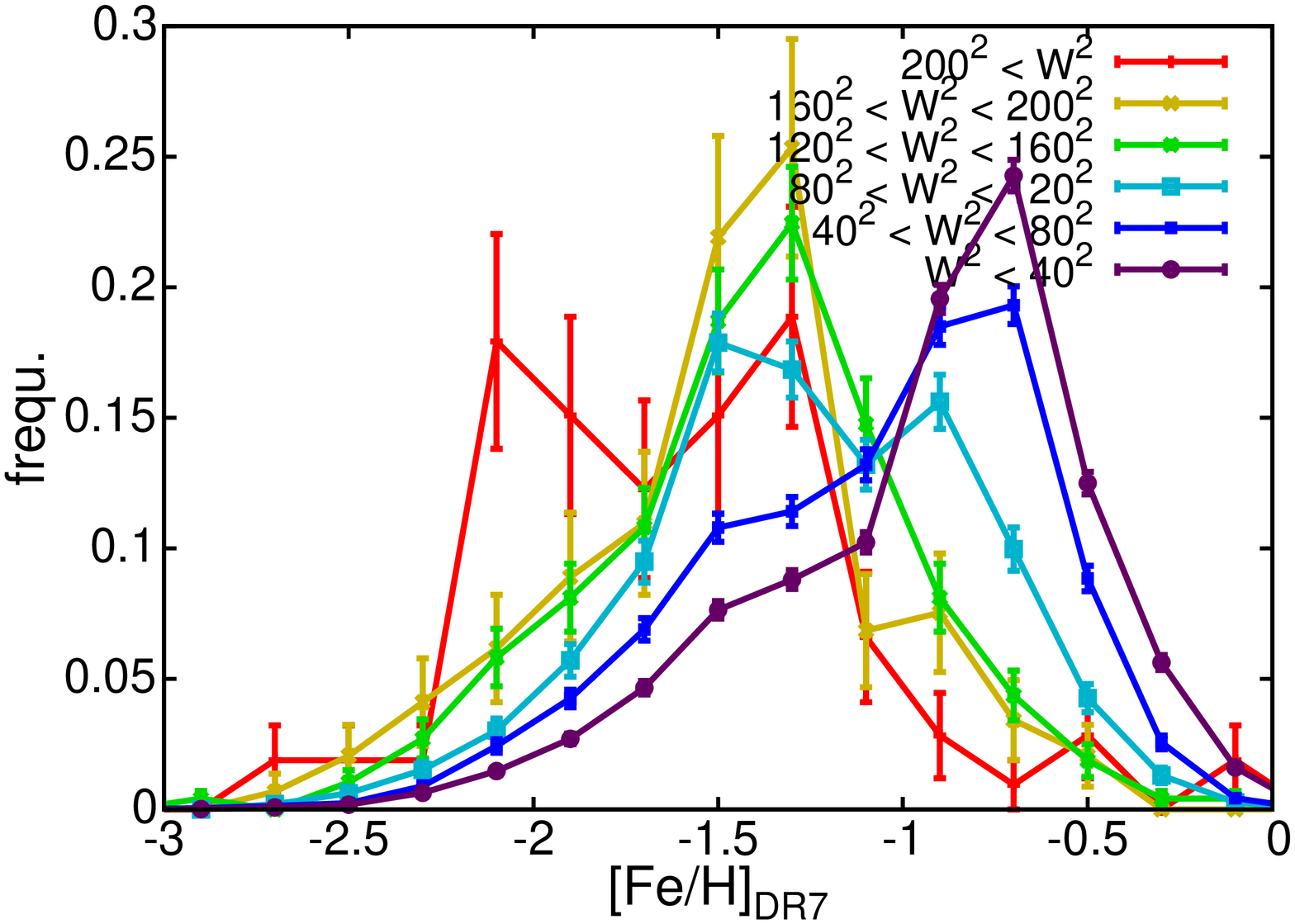,angle=-90,trim= 5mm 22mm 10mm 5mm,clip,width=0.325\hsize}&
\epsfig{file=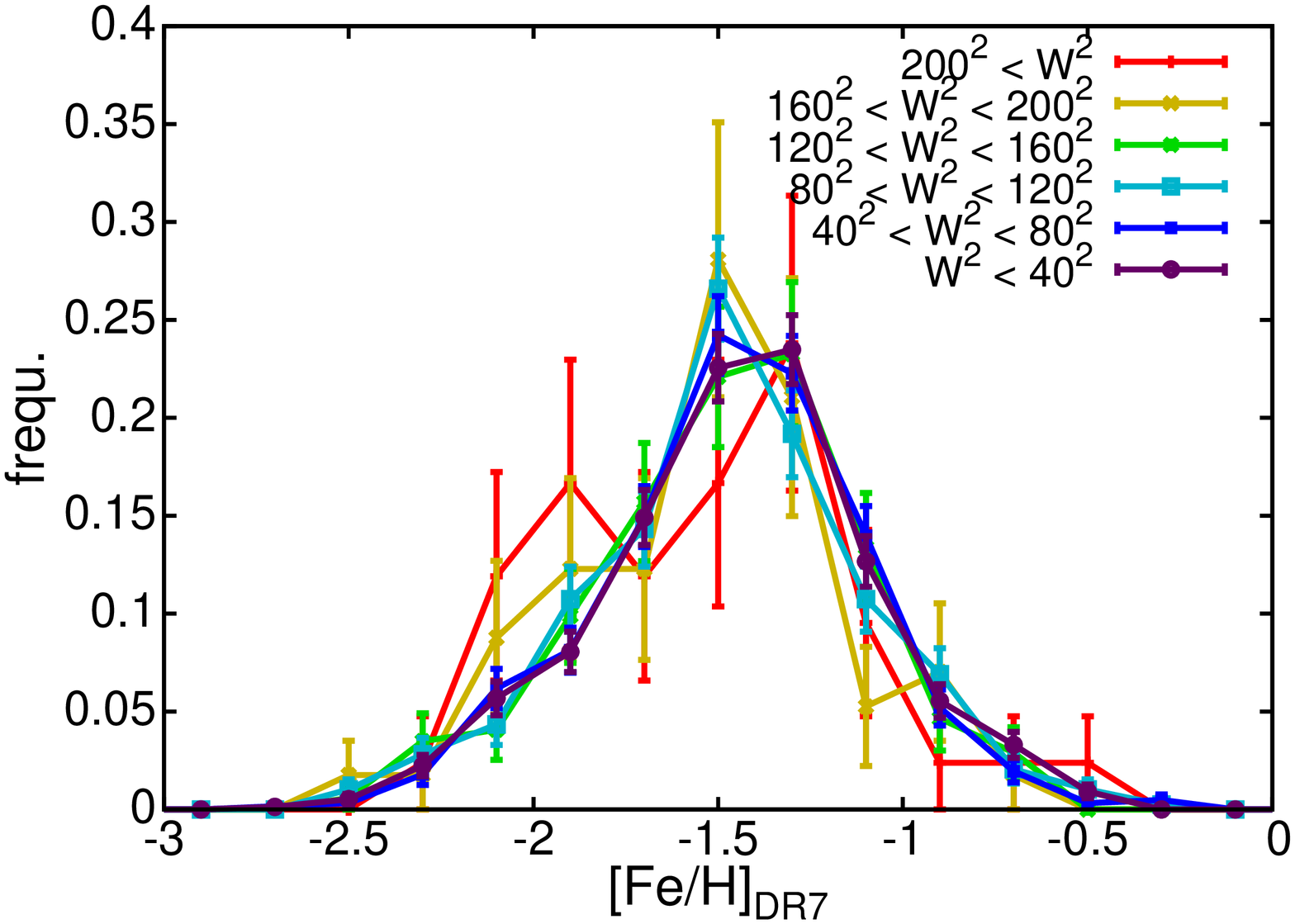,angle=-90,trim= 5mm 22mm 10mm 5mm,clip,width=0.325\hsize}\cr
\end{tabular}
\caption{Metallicity distributions at different values for kinetic
  energy in the different distance prescriptions. From top to bottom we show the entire C10 sample, the dwarf stars from C10, the adopted main sequence calibration and the Ivezi{\'c} et al.(2008) age-dependent calibration. The left column displays a separation by the entire kinetic energy, while the centre column makes only use of the vertical velocity $W$. The right column shows the distributions using the vertical energy part for retrograde stars ($V < -10 \kms$) to reduce the disc contamination. Velocities are taken in $\kms$. Error bars indicate the Poisson errors.
}\label{fig:kinen}
\end{figure*}

Another way to look at the problem is to plot metallicity
distributions as a function of the kinetic energies (by stellar mass) of stars. In \figref{fig:kinen} we
show the separations in entire kinetic energy ($v^2$, left column) and in vertical
kinetic energy ($W^2$, middle column). In the rightmost column we show the vertical kinetic energy, but for retrograde stars ($V < -10 \kms$). As the rotation velocity of disc
stars adds to the kinetic energy the disc has its strongest dominance
of course not in the lowest energy bin, but at $180^2 (\kms)^2 < v^2
< 270^2 (\kms)^{2}$, as most disc stars are on quite circular orbits with an entire velocity close to the circular rotation speed of the Galaxy. The prominent shift especially in the left tail and also peak of the metallicity
distributions in the original C10 distance prescription
for the full sample (top row) can be seen in both the total and vertical energy separation. However,
this already diminishes, when we
plot the subsample of the C10 dwarf stars with $\llg > 4.0$ (second row). No convincing outer halo signature can be found in either of the native SDSS calibrations (bottom row). At the highest vertical energies a single spike at $\feh \sim -2.2$ sticks out. At first glance this could be taken as a hint for an outer halo. However, in this
interpretation it should be mirrored by the entire energy
distribution, which is not the case. We are tempted to identify this
feature at least partly with a prominent metal-poor stellar stream described by \cite{Helmi99} and \cite{Kepley07}, which we expect exactly at these very high vertical
energies. This stream can as well serve as explanation for the subtle and insignificant
increase of vertical velocity dispersions at $\feha < -1.8$
that was seen in \figref{fig:vertkin}. 

There appears also a slight general drift towards lower metallicity at the highest energies, which can be mainly seen in the vertical term. Caution should be exercised, however, due to the varying presence of the Galactic disc (most prominent at the lowest vertical kinetic energies and at total energies corresponding to the rotation speed), which then impacts on the normalization of the halo component and fools the eye because the apparent halo peak can be shifted by the wing of the disc distribution. A quite robust approach is removing all prograde stars from the sample to minimize disc contamination. As this biases kinematics, the overall energy distributions are altered, but the vertical energies (right hand column) should not be affected.  In \figref{fig:kinen}, the entire C10 sample shows a clear signature of lower metallicities in the higher vertical energy bins. When removing the contested turn-off stars (i.e. plotting dwarfs, centre row) no trend apart from the discussed spike can be detected regardless of the applied distance determination. 

In summary it can be stated that also in terms of metallicity versus energy the sample has no reliable outer halo signature, neither when using the native SDSS calibrations nor when using the C10 dwarf subsample.

\begin{figure*}
\epsfig{file=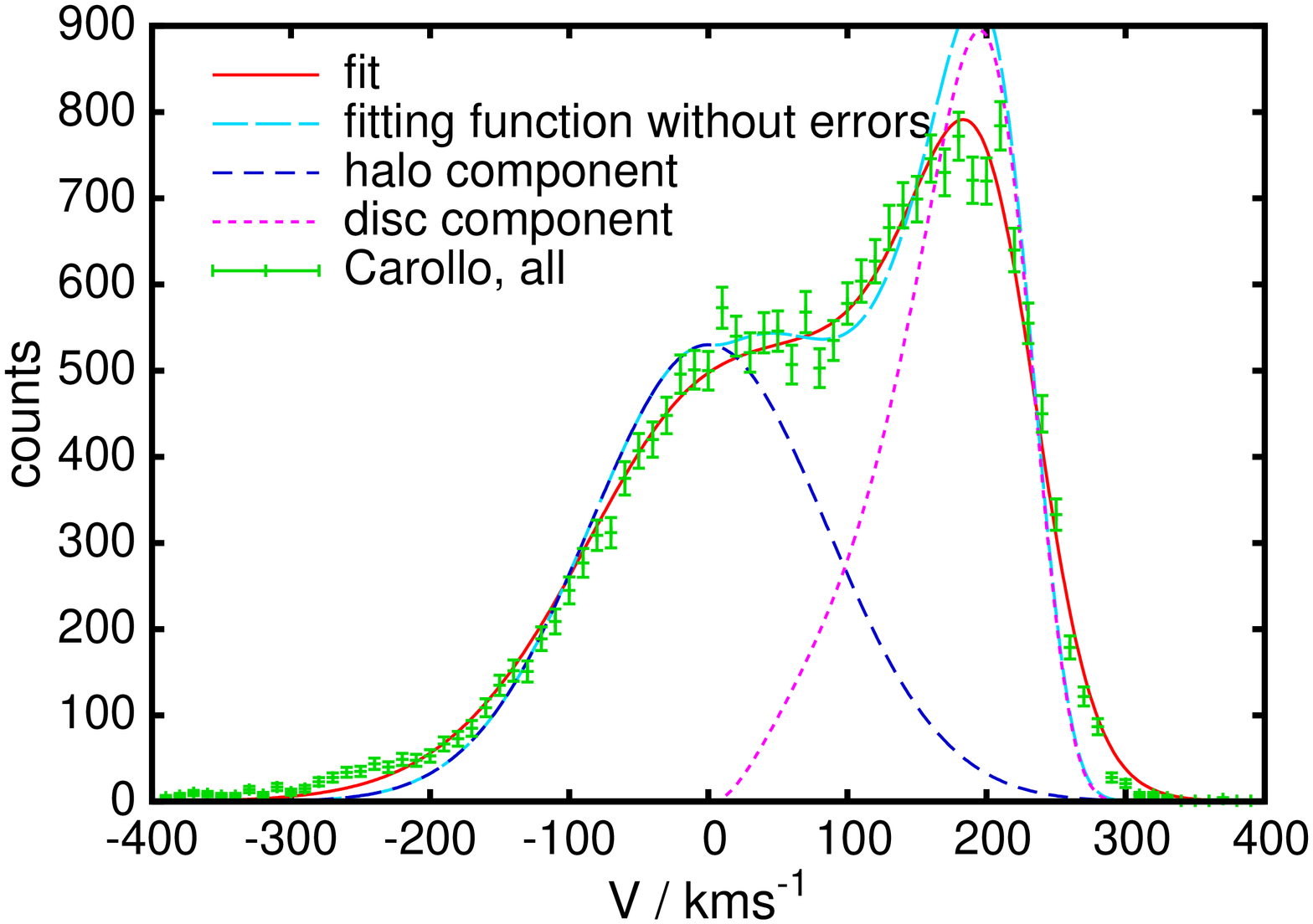,angle=-90,trim=5mm 15mm 5mm 3mm,clip,width=0.4\hsize}
\epsfig{file=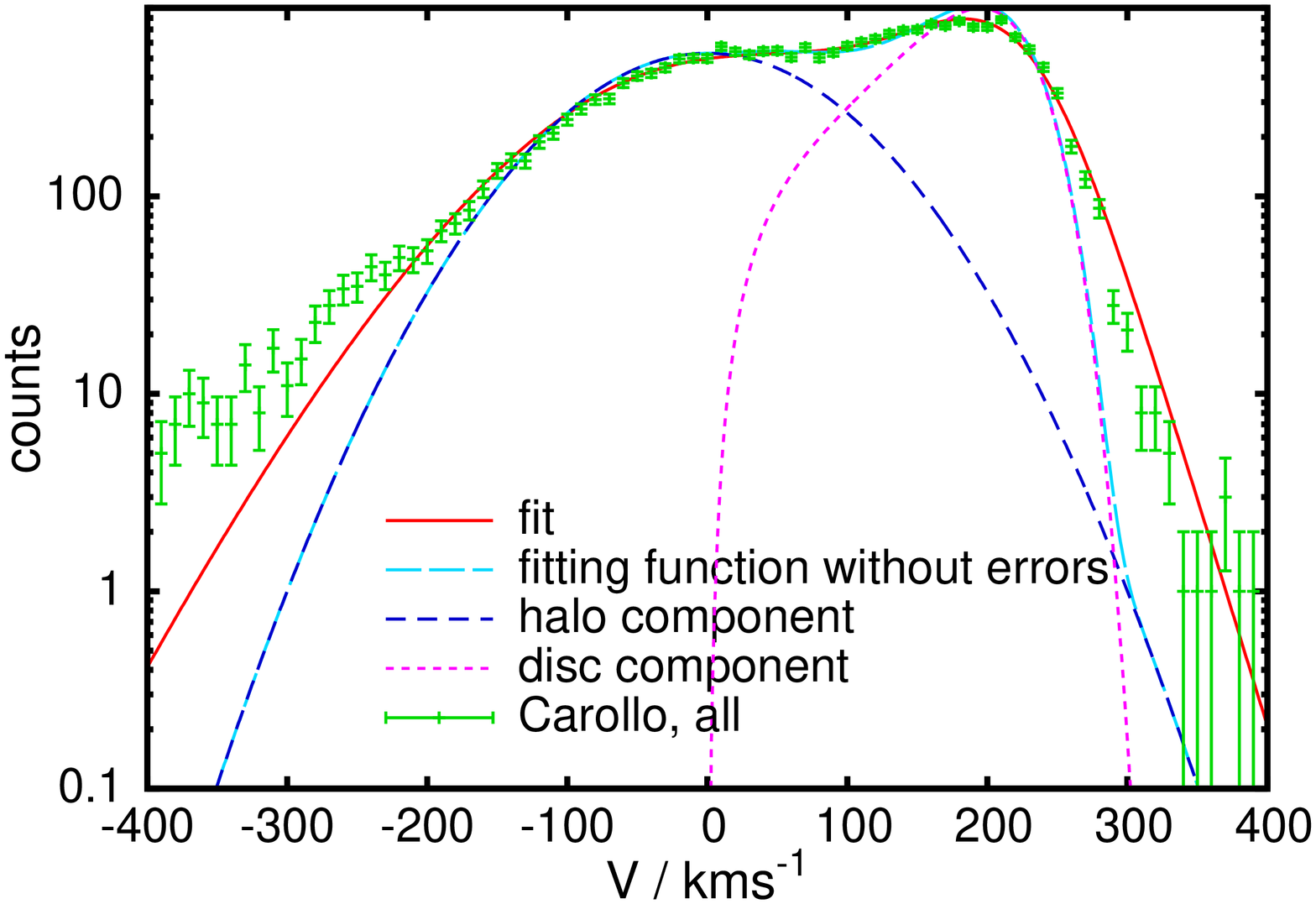,angle=-90,trim=5mm 15mm 5mm 3mm,clip,width=0.4\hsize}
\epsfig{file=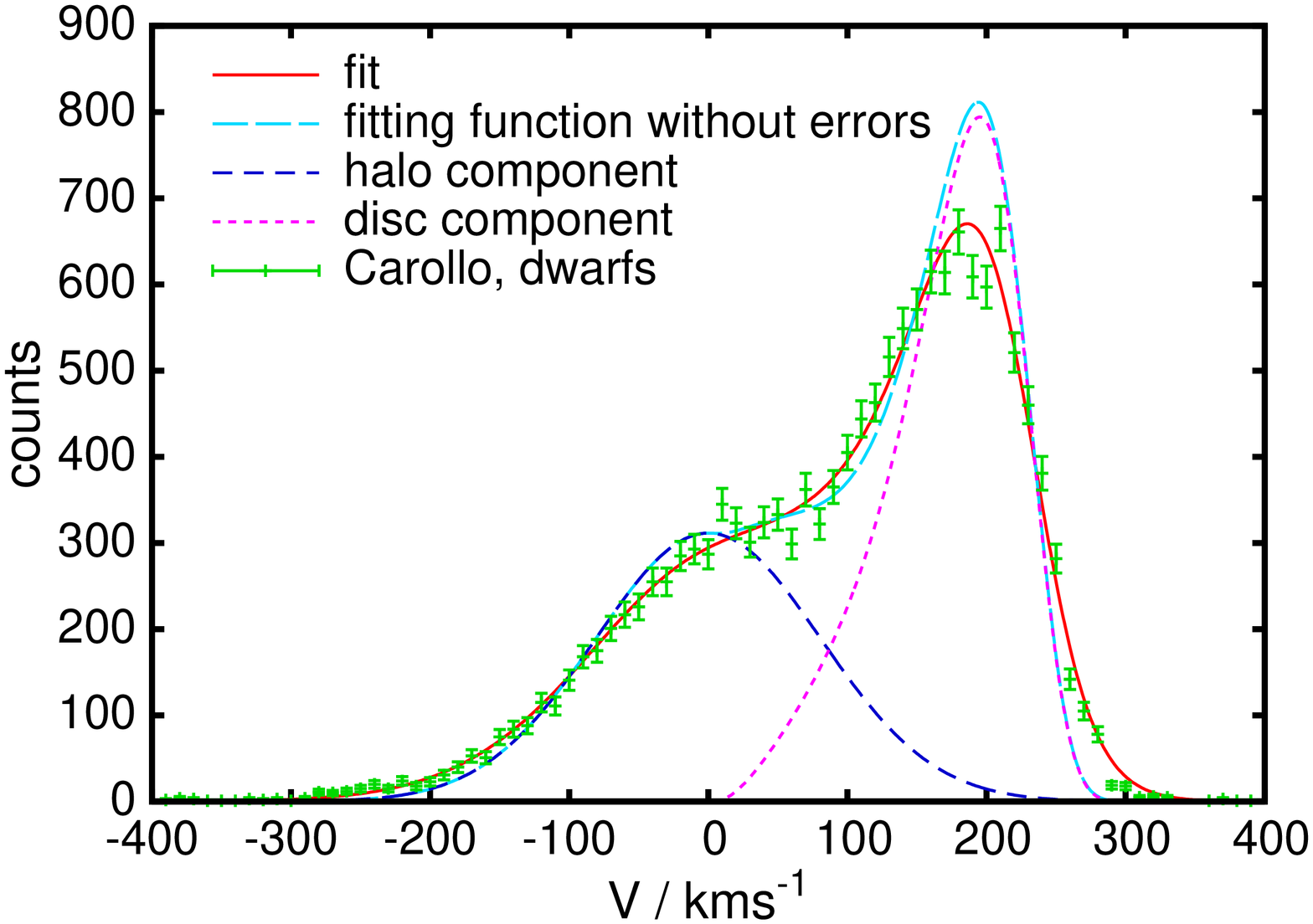,angle=-90,trim=5mm 15mm 5mm 3mm,clip,width=0.4\hsize}
\epsfig{file=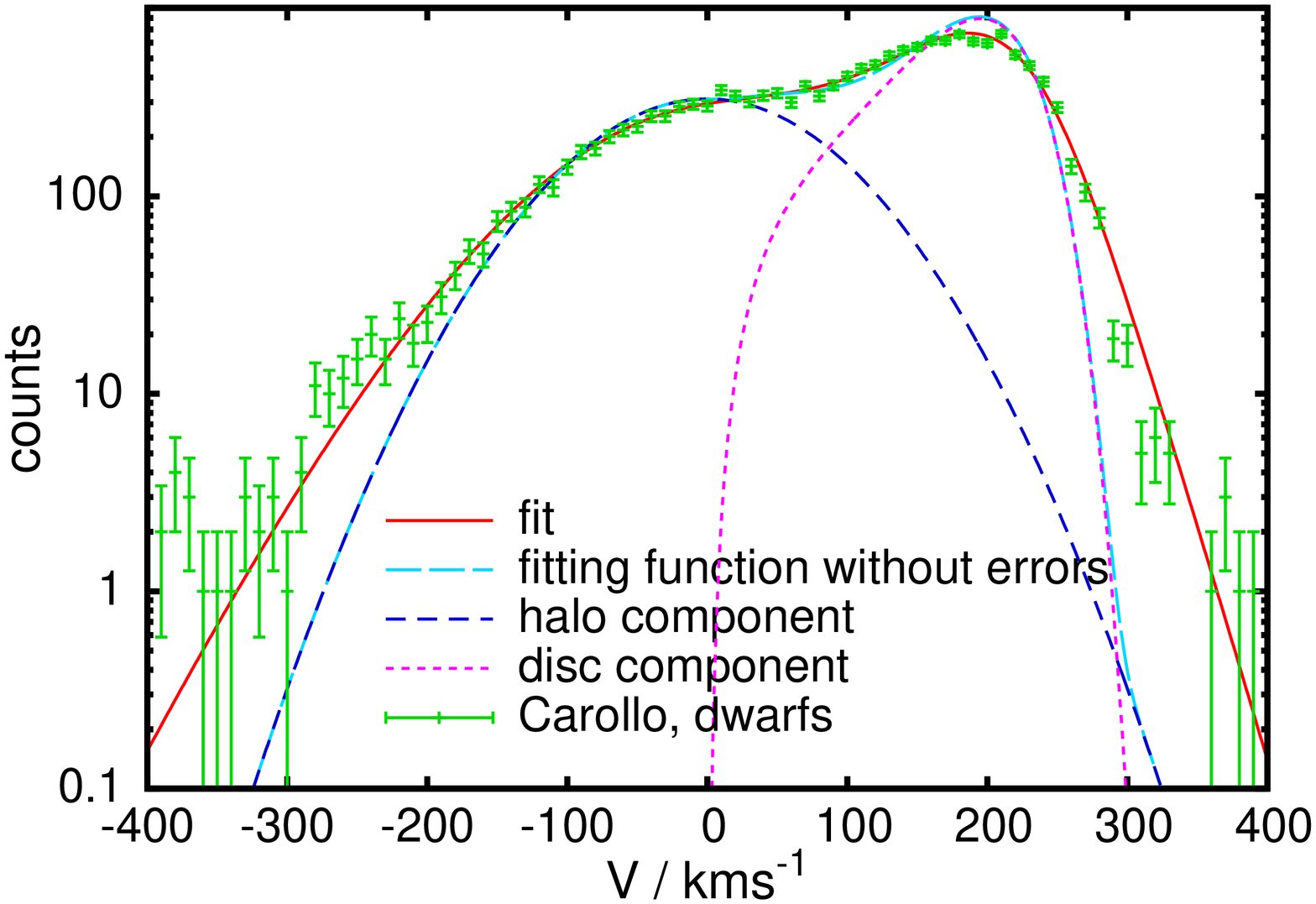,angle=-90,trim=5mm 15mm 5mm 3mm,clip,width=0.4\hsize}
\epsfig{file=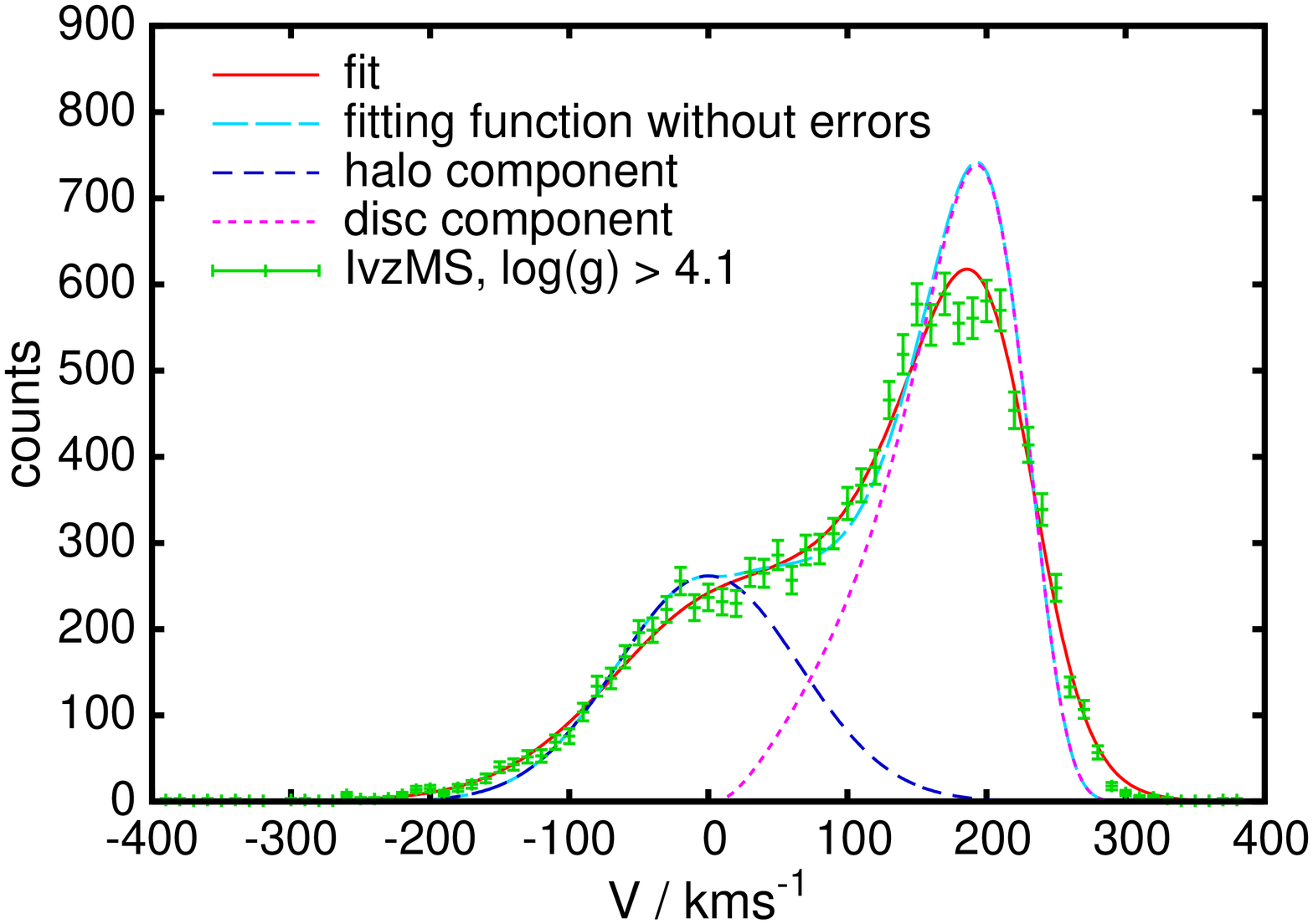,angle=-90,trim=5mm 15mm 5mm 3mm,clip,width=0.4\hsize}
\epsfig{file=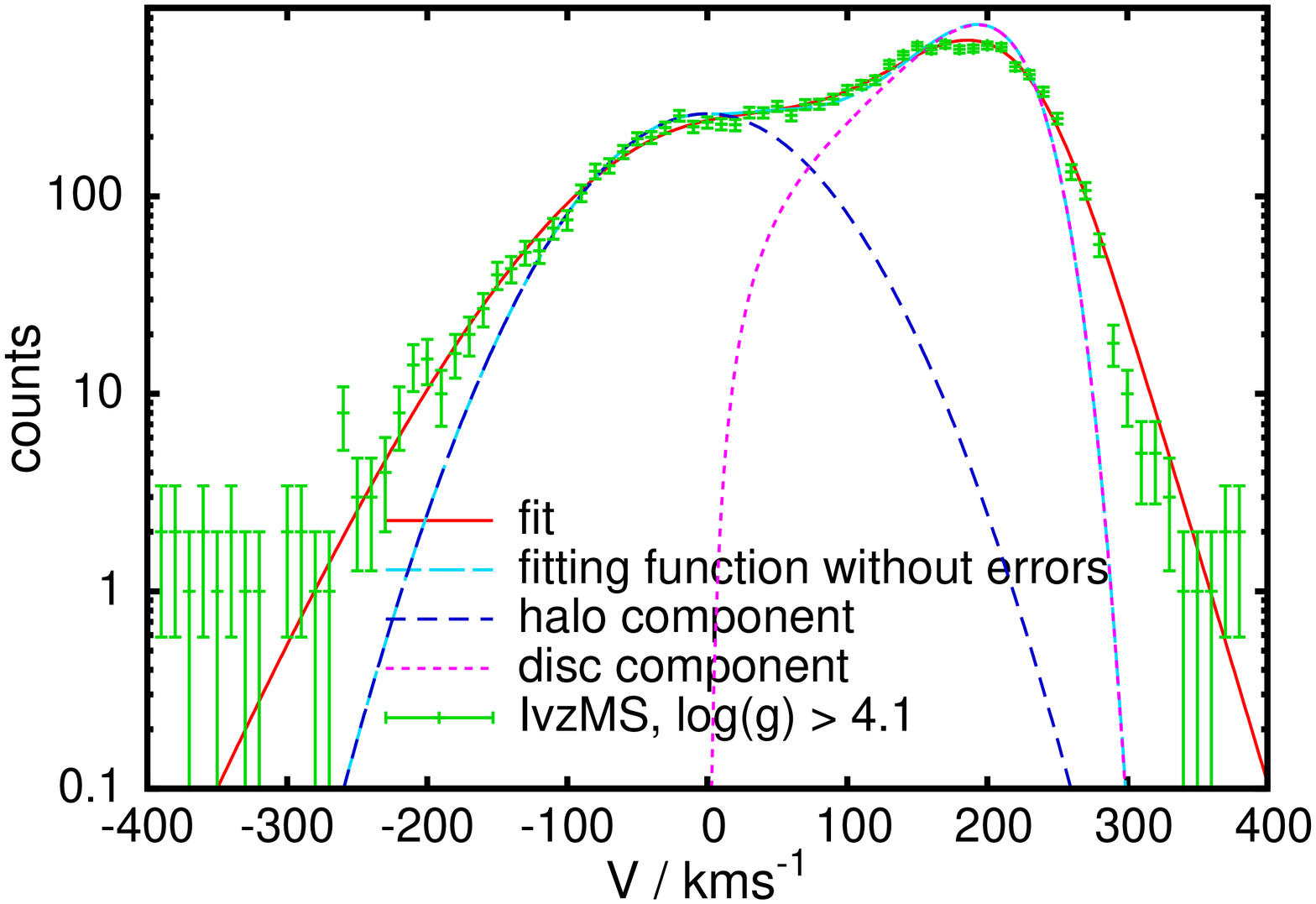,angle=-90,trim=5mm 15mm 5mm 3mm,clip,width=0.4\hsize}
\epsfig{file=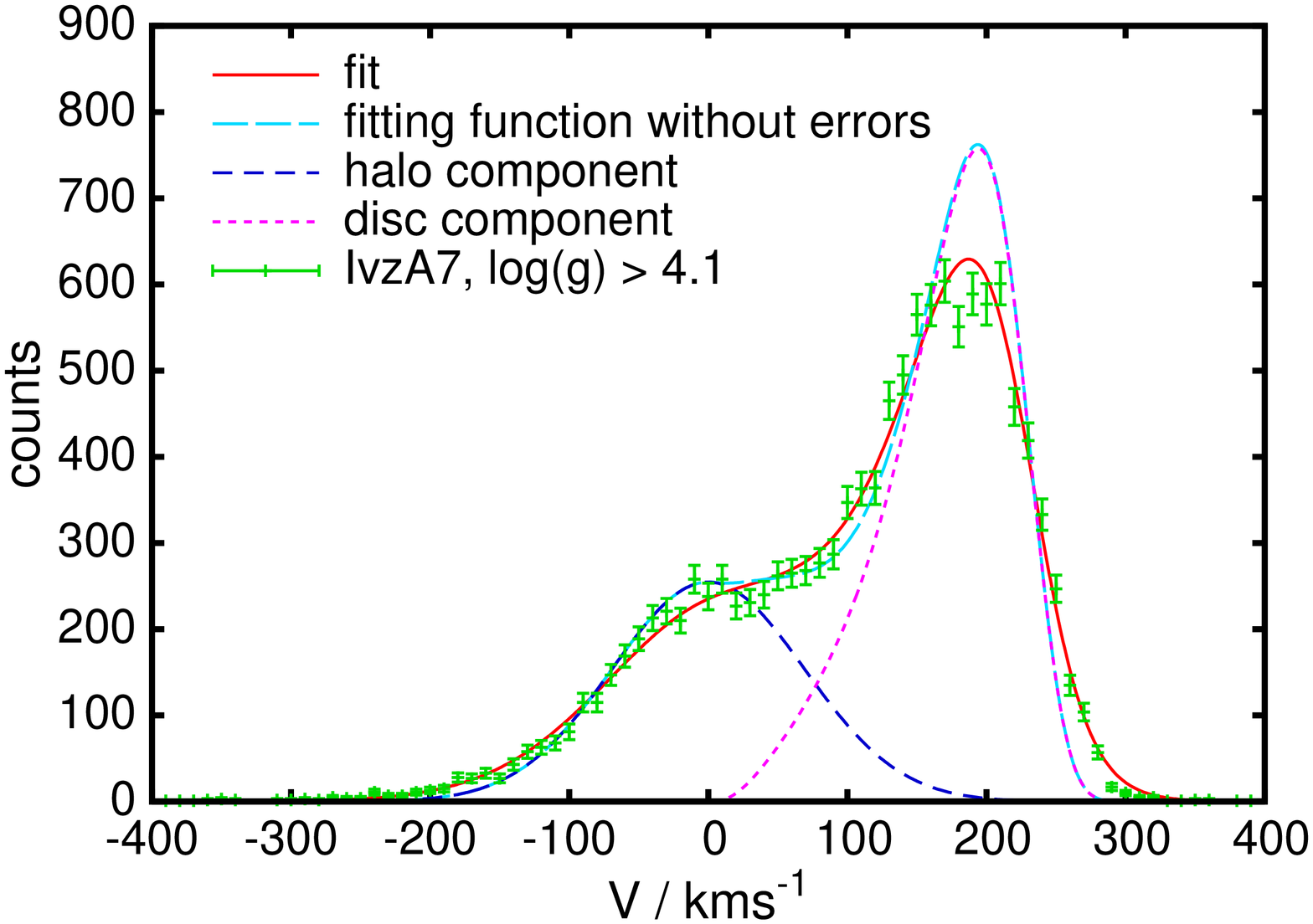,angle=-90,trim=5mm 15mm 5mm 3mm,clip,width=0.4\hsize}
\epsfig{file=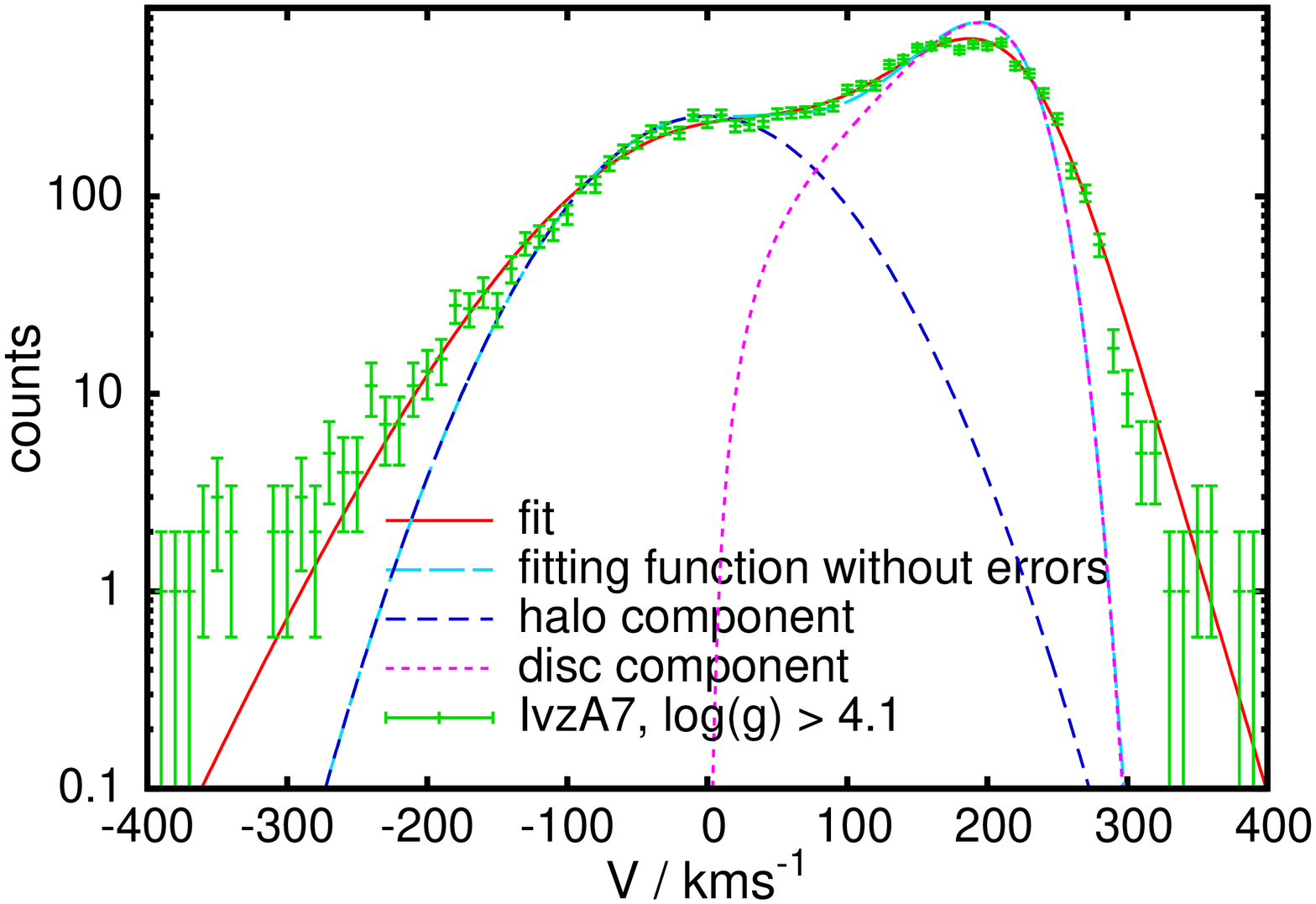,angle=-90,trim=5mm 15mm 5mm 3mm,clip,width=0.4\hsize}
\caption{Velocity distributions for stars (light blue, with Poisson
  errors) in the calibration sample for different selections and distance prescriptions: From top to bottom: The entire C10 sample, the C10 dwarf stars with $\llg > 4.0$, the dwarfs with $\llg > 4.1$ using the adopted main sequence calibration distances and the age-dependent formula from Ivezi{\' c} {\u Z} et al. (2008). To show both distribution centre and wings we contrast the linear scale (left) with logarithmic plots of the distributions (right). All distributions are 
fitted by a simple model (red line) with a Gaussian halo and a single (non-Gaussian) disc component. We did an error propagation from the values given in DR7 which is -- together with a magnitude error of $0.25 \mag$ - folded onto the original distribution (light blue). The error broadens the peaks and thus lowers the maximum count rates especially of the disc peak. The dark blue line shows the underlying Gaussian halo component and the purple curve the analytic disc component.}\label{fig:velos2}
\end{figure*}

\subsection{Component fit}\label{sec:comp}

\figref{fig:velos2} shows fits to the azimuthal velocity distribution (green data points) for the the four different subsamples and distance calibrations we use. Both columns show the same data, the left column on a linear scale, the right column on a logarithmic scale to examine both centres and wings of the distributions. 
We apply a simple Gaussian halo and a single non-Gaussian disc component. 
All fits were done for $-250 \kms < V < 280 \kms$, a limit that stays out of the regions dominated by noise, but reaches on the retrograde side still well into the region where any suspected retrograde halo component would be influential. To account for the observational errors, we folded the underlying distributions with a sum of Gaussians (with a spacing of $1 \kms$), of which the relative weights were derived globally for each component from error propagation on the single objects (stars with $V > 50 \kms$ attributed to the disc, objects with $V < 10 \kms$ to the halo) concerning proper motion and radial velocity errors. This treatment is a bit crude and we suspect that it underestimates errors on the left wing of the distribution and overestimates them on the right wing, which we identify as the most likely reason for the overshooting of the model against the data at $V > 270 \kms$. In a second step we folded with another Gaussian magnitude error of $0.25 \mag$ to account for uncertainties in the intrinsic brightness and thus distance of the stars and thereafter a third Gaussian of $7 \kms$ to account for velocity crossovers via the distance uncertainties.

For the disc component we make use of the analytic formula of Sch\"onrich \& Binney (in prep.). The underlying assumption is that of an isothermal sheet through the Galaxy with increasing velocity dispersions towards the centre, i.e. the stellar populations at each galactocentric radius are given a specific vertical and horizontal energy dispersion. For those populations their likelihood to be in the solar annulus and their local scaleheight can be estimated assuming a simple potential with constant rotation speed. We used a solar galactocentric radius $R_0 = 8.0 \kpc$, a circular rotation speed of $v_c =  220 \kms$, a disc scalelength of $R_d = 2.5 \kpc$, a scalelength for
 the vertical dispersion of $R_{\sigma_z} = 5.0 \kpc$ and for the
 horizontal dispersion as $R_{\sigma} = 7.5 \kpc$. The adiabatic 
 correction index (index dependent on the shape of the potential that describes the change of vertical energy along orbits that are extended over different galactocentric radii) was set to 0.5. As we are using the adiabatic correction without a recorrection for energy conservation, the disc local scale height gets moderately underestimated, an effect that we partially cope with by setting the adiabatic correction index to $0.5$, slightly below the expected value for the upper disc. To simplify the calculation we summed up disc contributions at equal parts at altitudes $z = 400, 700, 900, 1100, 1400, 1800, 2400 \pc$. The Gaussian halo component was set at rest ($V = 0$). The five free parameters of the fit were disc and halo normalization ($N_h = 530, N_d = 16691$ for the entire C10 sample, $N_d = 14403, N_h = 312$ for the C10 dwarfs, $N_d = 13031, N_h = 262$ for the adopted main sequence calibration, $N_d = 13254, N_h = 255$ for the age-dependent calibration), halo azimuthal velocity dispersion ($84.7, 80.9, 65.4, 68.9 \kms$), disc local horizontal dispersion ($\sigma_0 = 43.2, 41.4, 41.5, 40.1 \kms$), and local scaleheight ($h_0 = 571, 499, 466, 439 \kms$). As seen in \figref{fig:velos2}, apart from a weakness on the high velocity side, which presumably derives from the use of a global observational scatter that could overestimate the uncertainty for disc stars, the two-component approximation gives decent fits (red lines) to the velocity distributions for the dwarf samples. 
The relative halo normalization varies between the different datasets as the most metal-poor stars are more likely to be classified into the turn-off or subgiant bands. 

Of most interest for this discussion is the shape of the (retrograde) halo velocity distribution. For the entire C10 distance sample (left column) we can clearly identify the bump that encouraged C07 and C10 to fit a separate velocity peak starting from around $-200 \kms$. This anomaly becomes especially apparent in the logarithmic plot and is confirmed by Poisson loglikelihood values ruling out equality of the theoretical (simple Gaussian halo) and observed distributions at highest significance. 

We point out that even this strong anomaly in the fit alone would not be a sufficient justification for a second physical component, especially not for a retrograde component. There is no reason to firmly believe that the Milky Way halo or its possible components should have a strictly Gaussian velocity structure; the disc certainly can not be adequately described by Gaussian fits \citep[a discussion of this can be found in][]{Strom27} and similarly the halo might have a more complex velocity distribution. The only way to cleanly identify a retrograde velocity distortion on the Galactic halo from kinematics would be to prove a difference to the prograde halo tail. As this is, however, impossible at the required accuracy due to the disc contamination, we conclude that without a clean halo-disc separation any attempt to use the azimuthal velocity distribution for proving a retrograde halo component is unreliable.

Besides its doubtful significance for indicating a separate component, the extended retrograde tail anyway diminishes when we turn to the dwarf samples, which do not contain the at least partly unphysical turn-off stars: restricting the C10 sample to their dwarf stars (centre column of \figref{fig:velos2}) the bump below $-200 \kms$ disappears. There is a weak surplus of stars between $-220 \kms$ and $-280 \kms$, which could, however, just be a statistical fluctuation. This is confirmed by statistics, which have the p-level (measured under the assumption of Poisson errors) for equality of the fit and the measured distribution at $0.55$ for $-250 \kms \le V \le 20 \kms$ (and $0.20$ when we extend the range to $-300 \kms \le V \le 20 \kms$) meaning that when drawing realisations from the given theoretical distribution, about $55 \%$ of the samples would be more different from the theoretical distribution than the current one. Regarding our imperfect treatment of distance errors the quality of the fit is rather surprising.
Using the main sequence distance calibration the distribution gets even more contracted due to the shorter distances. Apart from the slight surplus of stars left of $V \sim -320 \kms$, which just appear to be some sample contamination (indeed a distance test on the $100$ objects with lowest $V$ velocities still reveals a $2 \sigma$ distance overestimate) and the single bin with $8$ stars around $V = -260 \kms$ no anomaly is traceable any more, confirmed by statistics. Using the age-dependent calibration from \cite{Ivz08} the picture is similar. The fit looks even better down to $V \sim -300 \kms$ beyond which there is a weak surplus of about a dozen objects between $V = -400 \kms$ and $V = -300 \kms$. This can be attributed to misassignement of stars between the catalogues for proper motions and a probable non-Gaussian far tail of the proper motion error distribution.

It can also be argued that an attempt to fit the disc velocity distribution via a Gaussian fit increase the need for a spurious second halo component, especially as the large (unavoidable) velocity errors limit the information on the real shapes of the underlying velocity distributions: as the disc velocity distribution is naturally skewed towards lower velocities, a Gaussian fit will drop too steeply towards low rotation velocities. This invokes a second Gaussian for the disc (thin-thick discs), which will in most cases show the same problem again, forcing the halo component a little bit up into the prograde regime. This again increases the need for the creation of an artificial retrograde halo component. In this case we avoided this problem by using a more physical fitting formula for the Galactic disc, yet there was still noticeable influence by the disc fit onto the parameters of the simple Gaussian halo component.

Overall we can state that any striking excess of highly retrograde stars disappears from the distribution when we limit the sample to the more reliable dwarf stars. There is a good agreement between fits and data for the dwarf samples, apart from a misfit at high velocities and a slight surplus at the lowest velocities against a Gaussian fit, which, however, rather looks like imperfect error handling or a probable contamination of the sample with misidentifications in between the different catalogues delivering the proper motions or a non-Gaussianity in their contamination. Replacing the questionable Gaussian analysis for the disc by a more physically motivated formula and by applying an error propagation we have shown that neither a second halo nor a second or even third disc component are required for explaining the azimuthal velocity distribution.

\subsection{Velocity dispersions}

Using the adopted main sequence approximation and restricting the sample to $2020$ retrograde stars ($V < -10 \kms$) we can give a tentative estimate for the halo kinematics of $\sigma_U \sim 157 \pm 10 \kms$ and $\sigma_W \sim 75 \pm 8 \kms$. The latter is close to the azimuthal velocity dispersion of $\sigma_V \sim 70 \kms$ from the last section. 
In these values we accounted for the errors reported in the SEGUE pipeline plus a dispersion of roughly estimated $15\%$ in the distances. These act to reduce the derived velocity dispersions from the actually measured ones. Those values are considerably smaller in $V$ and $W$ than what was given by C10 partly due to the spurious crossover into $W$ velocities having diminished. We point out that these values agree well with the results of \cite{Kepley07} and \cite{Smith09}. Although we use the same sample with the same distance scale as \cite{Smith09} we cannot confirm their quoted errors of $2 \kms$, since the systematic uncertainties by the distances within the \cite{Ivz08} method and sample cleaning are currently too large to assess dispersions on a scale better than $5 \%$.

\section{Conclusions}

We have described how errors in distance estimates result in an apparent systematic retrograde motion of the Galactic halo, an effect to which the SEGUE/SDSS sample is especially prone by its strong poleward orientation. The general problem of distance biases similarly applies to any study that makes use of proper motion-based estimates.
We find that the distance derivation of \cite{Carollo07} and \cite{Carollo10} is flawed by sorting stars into unphysical positions in the HR diagram: objects are placed between the subgiant and dwarf sequences in positions that would require stellar ages in excess of the age of the universe. Despite the elegance of the general idea to sort stars into known sequences according to their estimated gravities, the method itself and the used gravity cuts are not well supported by measurements. Moreover there is no "turn-off"-sequence, but turn-off stars are populating a region that spans of order $1$ magnitude in luminosity. In this light the statement by C10 to have distances precise to about $10-20 \%$ is an unsupported claim.

From the distances kindly provided by Carollo et al. we calculated back to their assumed absolute magnitudes and found systematic differences of $\sim 0.2$ to $0.3 \mag$ and a large scatter for metal-poor main
sequence stars towards the adopted main sequence calibration as well as towards the age-dependent calibration by \cite{Ivz08}, also far to the red side of the suspected turn-off region. The adopted main sequence calibration is only slightly fainter than the theoretical BASTI isochrones.

We have shown in Section \ref{sec:HR} that the claim by C07 and C10 to have found a counter-rotating extended tail of the halo is largely caused by unphysical assumptions about locations of stars in the HR diagram, by magnitude uncertainties in the turn-off stars and by the use of a too bright main sequence calibration. The correctness of this tail can be ruled out by statistical tests, as described in Section \ref{sec:sign}. The tail diminishes when we limit the C10 sample to dwarf stars and disappears when we make use of the better founded \cite{Ivz08} calibrations, which are consistent with fiducials and isochrones. In Section \ref{sec:comp} we demonstrated that the halo distribution for the dwarf samples regardless of the applied distance determination can be fit by a simple Gaussian component. 

We have also shown that in the DR7 pipeline stars with lower metallicities are shifted towards lower gravities, considerably increasing their fraction among the thought-to-be turn-off stars. Further the magnitude difference for their main sequence stars against the \cite{Ivz08} main sequence calibration grows towards lower metallicities. The stronger prevalence of distance errors at the metal-poor end of the metallicity distribution will thus give any spurious counter-rotating tail members a biased metallicity distribution.

Finally we have shown that the claim of C10 that the counter-rotating component members reach to higher altitudes can as well be traced back to distance determinations: the dispersion of the vertical velocity component is
significantly increased by their distance errors, though due to the polewards sample orientation this effect would at first order be smaller than for the other velocities. As shown in Section (\ref{sec:sign}) simultaneously the $W$
velocities of the halo stars with distance overestimates are artificially increased by of
order $50 \kms$ via spurious velocity cross-over terms from the heliocentric azimuthal velocities in the
derivation. The effect is strongest for the most strongly retrograde objects (they have the largest heliocentric $V_h$) and is aggravated by their selection for stars in galactocentric radius $7 \kpc < R < 10 \kpc$.  This colludes with their metallicity dependent distance bias (see above) to produce their findings of decreasing metallicities at high altitudes. There is a slight excess of more metal-poor stars in a single velocity bin at high vertical velocities, which is not mirrored by the behaviour at high total kinetic energies. We argue that this is most likely a reflection of a well-known local stream that has been identified by \cite{Helmi99} and \cite{Kepley07}.

Another source of error is the modelling of especially the Galactic disc azimuthal velocity distribution by Gaussian components. It was shown by \cite{Strom27} that Gaussian modelling of the Galactic disc lead to unphysical results and the identification of spurious components on the low rotation side because of the extented tail. As the skewed $V$ velocity distribution enforces in most cases the introduction of a second Gaussian component, that - being a mere artifact by wrong assumptions - can then be misinterpreted as physical reality, Gaussian modelling of the Galactic disc in a combined disc and halo sample can wrongly force the halo component into the prograde regime to compensate for the two steeply falling disc terms. Consequently this then creates the need for inference of a retrograde component to compensate for the bias.

We also argue that magnitude based distance assessment schemes introduce a velocity bias that resembles the Lutz-Kelker bias: If the error in absolute magnitudes follows a Gaussian distribution, the distance error distribution will thus form an extended tail that grows stronger with increasing dispersions. Via the proper motion part in the determination of space velocities, which is proportional to the estimated distance, measured velocities develop extended tails away from the solar motion. For the $V$ velocity distribution of especially the halo this gives the halo an asymmetric velocity distribution with a longer tail in the retrograde regime, a process that can explain the moderate asymmetries found e.g. by \cite{Norris89}.

Finally we note that it is by no means imperative that the halo have a
Gaussian velocity distribution. In this light it is rather surprising that our
simple Gaussian halo component can fit the data so well. Even if there were
deviations from Gaussianity this would alone be no convincing sign for a
separate component. A more convincing indication would be a proven difference
between the prograde and the retrograde tail of the halo azimuthal velocity
distribution, but disentangling this from disc contamination on the prograde
side will be difficult.

Recently Carollo and collaborators submitted a rebuttal paper \citep[][]{B11} claiming that our analysis presented here be wrong.
Instead of discussing all our arguments their revised analysis relies on
two central claims: They state that the distance scale adopted by us is
wrong and that there is an asymmetric halo azimuthal velocity distribution
for their metal-poor stars, neither of which we concur with.

Concerning the distance issue we stress that our conclusions are valid for
the Carollo et al. (2007, 2010) and both \cite{Ivz08} distance
calibrations; our work does not rely on a single distance
scale in contrast to claims made by \cite{B11}.
Beers et al. criticize us for adopting the incorrect main sequence
calibration of \cite{Ivz08} but failed to note that we actually
stretched this calibration in the same direction of their preferred one by
increasing the luminosities by $0.1$ magnitudes and accounting for
alpha-enhancement by increasing the measured metallicities by $0.2$ dex
(Sect. 2). Importantly, we have also made use of their preferred \cite{Ivz08} calibration (here denoted IvzA7) and find no significant
differences (e.g. Fig. 11). Finally, we have made use of direct isochrone
distances, which fully corroborates our findings.

As for the azimuthal velocity distribution of the most metal-poor stars,
we re-emphasize that any magnitude-based distance scheme invokes a bias in
the inferred distances and thus an asymmetric azimuthal velocity
distribution by definition; this effect is akin to the well-known
Lutz-Kelker (1973) bias and is illustrated in \figref{fig:velos2}. In view of a large magnitude scatter
(which their metal-poor stars clearly have, see their Fig. 5) a Gaussian fit is
inappropriate due to the missing error handling. In this light it is not
surprising that their new revised parameters are quite different from
their original results (e.g. for $z_{max} > 5 \kpc$ their „outer halo“ mean
velocity rose from $-128 \kms$ (cf. Table 1 in  C10) to $-59 \pm
20$ km/s). Finally, we argue that moving a considerable fraction of the wrongly identified
turnoff stars up to the subgiant/giant branch as done by Beers et al.
will make the distance overestimate for misidentified dwarfs among them even more
severe.

In summary our criticism of the C07, C10 works remains in full: our in-depth re-analysis of their data with different
distance calibrations and a proper error handling reveal no convincing
evidence for a dual halo.

The systematic distance uncertainties make it dangerous to draw a definitive conclusion for the strength or existence of a possible counter-rotating halo component. All current distance calibrations have problems and need improvement before a method along the lines used by C07 and C10 can be attempted. We would like to stress that we do not and would not want to rely on either of them. Two central conclusions can, however, be drawn without having to trust any of the different distance calibrations: Even on the C10 or C07 sample using their distances, no reliable detection of any non-Gaussianity in the halo, be it a counter-rotating halo or not, is possible on the examined data set in any of the dwarf samples. If a separate component gets detected on a larger sample in the future, it should be significantly weaker than what was claimed by C10.

\section{Acknowledgements}
It is a pleasure to thank Heather Morrison and James Binney for fruitful discussions and helpful comments on the draft.
We thank Tim Beers for kind provision of the C10 data and helpful answers to our questions. R.S. acknowledges financial and material support from Max-Planck-Gesellschaft and Max Planck Institute for Astrophysics.

\section{Appendix}

\subsection{Kinematics and geometry}\label{sec:geo}

We use up-to-date values for the basic constants
of our Galaxy. 
The rotation speed is assumed to be $220 \kms$ in concordance with
recent results of \cite{Koposov10}. We apply the recent determinations of solar
motion relative to the local standard of rest from \cite{SBD}, which
is $\vsun=(11.1_{-0.75}^{+0.69}, 12.24_{-0.47}^{+0.47}, 
7.25_{-0.36}^{+0.37})\kms$ with additional systematic uncertainties
of $\sim(1,2,0.5)\kms$. Neither the Galactic rotation rate nor the solar galactocentric radius $\Rsun$ are very well determined,
but the angular motion of Sagittarius $A^*$ is \citep[][]{ReidBrunth04}. The galactocentric radius of $\Rsun = 8.5 \kpc$ assumed in
C10 is inconsistent with their assumption for the rotation speed. We adopt $\Rsun = 8.0 \kpc$, which is in
concordance with most measurements, and coincides with the
most recent trigonometric parallax determination for Sagittarius B2 \citep{ReidB209}.

\subsection{Distance calibrations and metallicities}

For adopted distances in SDSS colours two alternatives exist:
The \cite{Ivz08} calibration (see Appendix therein) or the Beers
(2000) calibration. A third possibility would be using the isochrones
directly, which would allow for a statistical
implementation of the subgiants and also allow for natural shape changes: however, their handling is beyond the scope of this work. The \cite{Ivz08} main sequence calibration uses the formula:
\begin{equation}
M_r(g-r, \feh) = 
\end{equation}
\begin{equation}
= 1.65 +6.29(g-i)_0-2.30(g-i)_0^2 -1.11\feh -0.18 {\feh}^2
\end{equation}
where $M_r$ is the adopted absolute magnitude and $(g-i)_0$ is the dereddened SDSS
colour index. A short assessment of this formula reveals that it fits the zero age
main sequence relatively well in the required colour range at low
metallicity. Apart from the shortcome that the subgiants are not considered, the metallicity dependence is not well matched at the high metallicity end. Further the colour dependence is not perfect at the level of precision required for
distances and there are no crossterms between metallicity and colour,
i.e. there is no implementation of metallicity-dependent shape
changes. Thus this relation has to be used with caution. A comparison with the BASTI isochrones reveals that the \cite{Ivz08} calibration at low metallicities is slightly fainter than the isochrones without alpha enhancement. To correct for alpha enhancement we slightly increase the metallicity of all metal-poor stars (with $\feha < -1.0$) by $0.2 \dex$ and let this correction go linearly to zero between $\feha = -1.0$ and $\feha = 0$. In addition we increase all luminosities by $0.1 \mag$ to reconcile the calibration better with the main sequence of $M92$ according to the fiducials by \cite{An08}, to avoid any suspicion to have buried the counter-rotating tail by a too faint calibration, and as we detected a minor distance underestimate in our statistics without this shift. Further the calibration is now just slightly brighter than the isochrones as
seen in \figref{fig:HRrgi}, which is a desired effect at those metallicities, as there are indications that metal-poor isochrones could underestimate the luminosities (or equivalently overestimate
the effective temperatures) of lower main sequence stars \citep[][]{Casagrande07}.

To account for age effects \cite{Ivz08} also suggested an "age-dependent" calibration, which rises more steeply towards the blue side (Eq. $A7$ in their appendix). To meet concerns of our referee we show all relevant statistics also in the light of this other calibration. This calibration is steeper in colour than the main sequence calibration, i.e. it is brighter at blue colours and fainter at red colours, following an intention to cope with the steepening of isochrones near the turn-off. However, the steepening of isochrones only applies near the turn-off, while this relation is globally inclined against the main sequence of the isochrones as well as against the main sequence calibration. As this relation thus does not follow the blueward shift of the turn-off towards lower metallicity, a relative overestimate is expected for the distances of the blue stars at lowest metallicity. Vice versa the distances to metal rich turn-off stars might be underestimated.

We tested all our results for the C10 dwarf sample and also for cutting in colour and different gravity selections to delineate the impact by the turn-off region. As can be seen from the central panel in \figref{fig:HRrgi} the difference between the adopted main sequence calibration and the C10 magnitudes persists also to the red side of the turn-off region and is also present for the \cite{Ivz08} age-dependent relation.

\begin{figure}
\epsfig{file=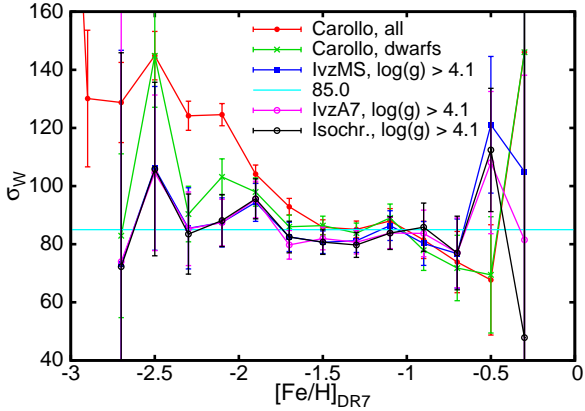,angle=-90,width=\hsize}
\caption{Vertical dispersions vs. metallicity revisited. The lines are the same as in \figref{fig:vertkin}. Additionally we plot in black the same derived via direct isochrone determination of distances.}\label{fig:vertkiniso}
\end{figure}

As an even fourth distance determination we made use of the BASTI isochrones \citep[][]{Piet04, Piet06}. To accomplish that we account again for alpha enhancement with the same prescriptions as for our adopted main sequence relation. For each star we choose the closest $12.5 \Gyr$ isochrone in metallicity from a dense grid kindly provided by S. Cassisi for our age determinations in \cite{CSA11}. On this isochrone we choose the r-band magnitude from (g-i) on the main sequence. When the turn-off is to the red of the stellar position, we extrapolate from the turn-off point of the isochrone following the shape of the \cite{Ivz08} main sequence relation. So these stars are placed at an extrapolated turn-off point. As one can easily see from \figref{fig:vertkiniso} in the central statistics there are no significant changes towards the other main sequence calibrations. We confirmed that no real difference was detectable.

C10 used a slightly different metallicity scale from DR7, which was reasoned
to balance out a possible metallicity overestimate on the lowest
metallicity end: 
\begin{equation}\label{eq:metdev}
\fehC = -0.186+0.765\feha - 0.068{\feha}^2
\end{equation}
We argue that their claimed overestimate at low metallicities most
likely reflects the impossibility of the pipeline to cope with the
faint metal lines in this region, getting lost in the low resolution
and low signal-to-noise ratios. This effect is quite similar to the
loss of accuracy e.g. in Str\"omgren photometric metallicities. The
formula has the dissatisfactory effect of aggravating the
well-known bias of the DR7 pipeline to underestimate the metal content
of metal-rich stars. This can be seen by comparison of local stars from the Geneva-Copenhagen Survey \citep[][]{Nordstroem04} with high vertical energy, to the
metallicity distribution of the SEGUE disc stars. Similarly the ugriz
metallicity calibration fails to reproduce metallicities
already slightly below solar metallicity as can be seen from Fig. $11$
in \cite{Arnadottir10}. The latter
problem is not of major importance for the halo, but applying the
formula by Carollo et al. would exacerbate metallicity-induced errors
on the disc population. We estimate that distances are only weakly affected by this correction on the low metallicity side (i.e. mostly for halo stars) as the sensitivity of stellar
atmospheres to the logarithmic metallicity scale gets lower. This can
be seen in \cite{VandenBerg10} or for comparison in fig. $13$ of
\cite{Casagrande10}. 

We use the DR7 metllicities throughout the paper and do not apply the correction from \eqref{eq:metdev}, but checked that our findings are not significantly affected by switching the distance scale.

\subsection{Deriving space velocities}

The dataset contains information on magnitudes, colours, the distances, stellar parameters, radial velocities and proper
motions.
We can thus derive the velocities in the solar/local coordinate system by:
\begin{equation} \label{eq:vvh}
\begin{matrix}
U_h = - d \sin(l){\dot l} - d \cos(l) \sin(b) {\dot b} + \cos(b) \cos(l)
v_{\parallel}  \\
V_h = d \cos(l){\dot l} - d \sin(l) \sin(b) {\dot b} +
\cos(b)\sin(l)v_{\parallel}\\
W_h = d \cos(b) {\dot b} + \sin(b) v_{\parallel} 
\end{matrix}
\end{equation}
where $l, b$ are Galactic longitude and latitude, ${\dot l}, {\dot b}$ are the proper motion components in $l$ and $b$ and d is the assumed distance to the stars. Of special interest is the degree of support of the measurement
by radial motions, which are independent of distance biases and have
smaller errors. Neglecting the geometrical extension of the sample,
the quantity of interest is thus the term connecting the azimuthal
velocity in the heliocentric frame $V_h$ to the line-of-sight velocity $v_{\parallel}$:
\begin{equation} \label{eq:nuv}
\eta_V = \left| \cos(b) \sin(l) \right|
\end{equation} 
The term $\eta_V$ is $1$ where the $V_h$ velocity is measured directly
from the line-of-sight velocity $v_{\parallel}$ i.e. at $b = 0$ and $l = 90^{\circ}, 270^{\circ}$, its contours on the sample are depicted in \figref{fig:geo}.

Due to the extension of the sample to a radius of more than $3
\kpc$ projected on the Galactic plane a small angle approximation cannot be
reliably taken. Using the Galactic rest frame velocities in heliocentric coordinate $U_a = U_h+ \Usun, V_a = V_h  + V_G + \Vsun, W_a = W_h + \Wsun$ (where $V_G$ is the Galactic rotation speed, $\Vsun, \Usun$ and $\Wsun$ are
the components of solar motion relative to the local standard of rest) the correction from the local coordinate system can be
done via the Galactic angle 
\begin{equation}
\alpha = \arctan(\frac{d\sin(l)\cos(b)}{\Rsun - d\cos(l)\cos(b)})
\end{equation}
between the line sun-centre to the line star-centre:
\begin{equation}
\begin{matrix}
U = U_a\cos{\alpha} - V_a\sin{\alpha} \\
V = V_a \cos{\alpha} + U_a \sin{\alpha} - V_G \\
W = W_a
\end{matrix}
\end{equation}
Throughout the paper blank letters $U,V,W$ denote the corrected velocities
in a Galactic reference frame, which are useful for assessing kinematics,
$U_h, V_h, W_h$ denote the velocities in the rest frame of the Sun and heliocentric coordinates, which are the native setting for exploring kinematic biases. 

\subsection{Sample cleaning}
The DR7 calibration sample contains 42841 spectra. The steps for cleaning the sample and the subsequent reduction of numbers are listed in Table \ref{tab:cleaning}; subsamples used in this work and the plots are set in bold font.
However, a considerable fraction of these entries is
double, i.e. stars whose spectra have been taken several times, some
for observational reasons as to improve the signal to noise, some as
they were both listed among the photometric calibration stars and the
reddening calibration stars. Therefore it is necessary to clean DR7
samples by identifying any measurements that are within $1.5$ arcsec
of each other in both right ascension and declination or are within an angular distance of $3$ arcsec and have a $g$ band magnitude difference below $0.1 \mag$.  
We chose the entry that gives full information on proper motions and the latest measurement in case both contain it.
This leaves $33023$ unique calibration stars in our DR7 sample. 
The C10 sample was already cleaned by them and has the $4 \kpc$ distance cut (according to their distances) applied, but not the cut in galactocentric radius (which we do not apply either). Hence the number of unique stars in the C10 sample is only slightly reduced when we demand a cross-match by stellar position on our DR7 table for being less then $2.5$ arcsec apart both in right ascension and declination. There are some stars dropping out because of a missing cross-match and there were $52$ candidate double entries with identical position in the C10 sample. Among those objects some are $\sim 0.3 \mag$ fainter than their second entry and the corresponding entry in DR7. In total we found $41$ stars which are $\sim 0.3 \mag$ fainter in apparent magnitude than their counterparts in DR7. We checked that none of these differences has a significant impact on the results.

We also removed from the sample the stars with signal to noise ratio
$S/N < 10$ and those which are flagged by the SSPP for spectral abnormalities, for colour mismatches or for being a suspected or proven white dwarf, as well as those with particularly unreliable radial velocity
measurements or SSPP parameters. We thus cleaned the C10 sample from $\sim 1700$ flagged
stars. Superficial tests did not show any obvious problems caused by those stars. Following C10, we also excluded all stars without determined proper motions
and "clean'' the sample according to \cite{Munn04} requiring that both
$\sigra$ and $\sigdec < 350 \mas$,
with an additional requirement of those stars to have quoted errors in
each proper motion component of $< 5 \masyr$. We checked on the C10 sample that $224$ objects in their sample not passing this cut did not cause any noticeable biases. The full cleaning gives a final DR7
sample of $28844$ stars, which we will use when plotting "all" stars in DR7. In plots of the full C10 sample we use its counterpart of $21600$ stars that already fulfills at their distances the condition of having all distances smaller or equal to $4 \kpc$. 

\begin{table}
\begin{tabular} { l | c | r }
condition & DR7 & Carollo et al. (2010) \\ \hline
original sample entries & $42841$ & --- \\
cleaned sample by C10 & --- & 23647 \\
unique \& cross-matched & $33023$ & $23553$ \\
unflagged & $29655$ & $21828$ \\
$S/N > 10$ &  $29638$ & $21825$ \\
$4500 \le T_{\rm eff} \le 7000$ & $29601$ & $21823$ \\
$\feha, v_{\parallel}$ fine & $29584 $ & $21819$  \\
prop. motion fine & {\bf 28844$^a$} & $21600$\\ \hline
$\llg > 4.0$ & $17365$ & $15023$ \\
$\llg > 4.1$ & $13880$ & $12120$ \\
$d_{Ivz} < 4 \kpc, \llg > 4.0$ & $15808$ & $14763$ \\
$d_{Ivz} < 4 \kpc, \llg > 4.1$ & {\bf 12678$^b$} & $11894$ \\
$d_{Car} < 4 \kpc$ & --- & {\bf 21600$^c$} \\
$d_{Car} < 4 \kpc, \llg > 4.0$ & --- & {\bf 15023$^d$} \\ 
\end{tabular}
\caption{Numbers of stars at different cuts in the two samples. The quality cuts are applied cumulatively from the first row until the horizontal line. Below the horizontal line we show selected subsamples with gravity and distance cuts. The cuts down to the quality cut in proper motion are applied successively. Below the horizontal line we show the effects of differnt distance estimations and cuts on the number of remaining stars.\newline
Subsamples used in the paper: $^a$ "all star sample", $^b$ "Ivezic dwarfs", $^c$ Carollo all, $^d$ Carollo dwarfs.}\label{tab:cleaning}
\end{table}

\end{document}